\documentclass[12pt,preprint]{emulateapj}
\usepackage{amsmath}
\newcommand{\Ndata}{N_{\rm data}}
\newcommand{\Mmodel}{M_{\rm pix}}
\newcommand{\Thetae}{\Theta}

\newcommand{\sphi}{\sin{\phi}}
\newcommand{\cphi}{\cos{\phi}}
\newcommand{\szeta}{\sin{\zeta}}
\newcommand{\czeta}{\cos{\zeta}}
\newcommand{\stheta}{\sin{\theta}}
\newcommand{\ctheta}{\cos{\theta}}
\newcommand{\si}{\sin{i}}
\newcommand{\ci}{\cos{i}}
\newcommand{\est}{{\mathrm{est}}}
\newcommand{\mestl}{{\boldsymbol m}_{\est,\lambda}}

\newcommand{\eig}{\kappa}
\newcommand{\lambdaopt}{\ell} 
\newcommand{\lambdar}{\lambda}
\newcommand{\er}{{\boldsymbol e}_R}
\newcommand{\erd}{{\boldsymbol e}_R^\prime}
\newcommand{\elos}{{\boldsymbol e}_\mathrm{O}}
\newcommand{\es}{{\boldsymbol e}_\mathrm{S}}
\newcommand{\WI}{W_I}
\newcommand{\WV}{W_V}
\newcommand{\trans}{\mathrm{T}}
\newcommand{\dxi}{\xi^\prime}
\newcommand{\drho}{\rho^\prime}
\newcommand{\ldxi}{{(\log{\xi})^\prime}}
\newcommand{\lddxi}{{(\log{\xi})^{\prime \prime}}}
\newcommand{\ldrho}{{(\log{\rho})^\prime}}
\newcommand{\lddrho}{{(\log{\rho})^{\prime \prime}}}
\newcommand{\ldxit}{{(\log{\xi})^{\prime \, 2}}}
\newcommand{\ldrhot}{{(\log{\rho})^{\prime \, 2}}}
\newcommand{\curv}{c(\lambdar)}
\newcommand{\gv}{{\boldsymbol g}}
\newcommand{\dv}{{\boldsymbol d}}

\newcommand{\error}{\varepsilon}
\newcommand{\errorv}{{\boldsymbol \error}}
\newcommand{\tdv}{\tilde{\boldsymbol d}}
\newcommand{\Dm}{G}
\newcommand{\tG}{\tilde{G}}
\newcommand{\mv}{{\boldsymbol m}}
\newcommand{\mpr}{\hat{{\boldsymbol m}}}
\newcommand{\Cm}{{\boldsymbol C}_\mathrm{M}}
\newcommand{\Cd}{{\boldsymbol C}_\mathrm{D}}
\newcommand{\omegaorb}{\omega_\mathrm{orb}}
\newcommand{\omegaspin}{\omega_\mathrm{spin}}
\newcommand{\fc}{f_\mathrm{IV}}
\newcommand{\thetaobs}{\theta_\mathrm{O}}
\newcommand{\phiobs}{\phi_\mathrm{O}}
\newcommand{\appweight}{A}
\newcommand{\appcpdf}{B}
\newcommand{\appbayes}{C}
\newcommand{\applcurve}{D}
\newcommand{\appcurv}{E}

\shortauthors{Fujii and Kawahara}
\shorttitle{Mapping Earth Analogs}
\begin{document}
\title{Mapping Earth Analogs from Photometric Variability: \\Spin-Orbit Tomography for Planets in Inclined Orbits}
\author{Yuka Fujii\altaffilmark{1} and Hajime Kawahara\altaffilmark{2}} 
\altaffiltext{1}{Department of Physics, The University of Tokyo, Tokyo 113-0033, Japan}
\altaffiltext{2}{Department of Physics, Tokyo Metropolitan University, Hachioji, Tokyo 192-0397, Japan}
\email{yuka.fujii@utap.phys.s.u-tokyo.ac.jp}
\begin{abstract}
Aiming at obtaining detailed information of surface environment of Earth analogs, \citet{kawahara2011} proposed an inversion technique of annual scattered light curves named the spin-orbit tomography (SOT), which enables us to sketch a two-dimensional albedo map from annual variation of the disk-integrated scattered light, and demonstrated the method with a planet in a face-on orbit. 
We extend it to be applicable to general geometric configurations, including low-obliquity planets like the Earth in inclined orbits. 
We simulate light curves of the Earth in an inclined orbit in three photometric bands (0.4$-$0.5$\mu $m, 0.6$-$0.7$\mu $m, and 0.8$-$0.9$\mu $m) and show that the distribution of clouds, snow, and continents is retrieved with the aid of the SOT. 
We also demonstrate the SOT by applying to an upright Earth, a tidally locked Earth, and Earth analogs with ancient continental configurations. 
The inversion is model-independent in the sense that we do not assume specific albedo models when mapping the surface, and hence applicable in principle to any kind of inhomogeneity. 
This method can potentially serve as a unique tool to investigate the exohabitats/exoclimes of Earth analogs. 
\end{abstract}
\keywords{astrobiology -- Earth -- scattering -- techniques: photometric}

\section{Introduction}
\label{s:intro}

One of the ultimate goals in exoplanet study is the discovery of extraterrestrial life. 
Recently, the abundance of Earth-sized exoplanets has been implied from the radial velocity observations \citep[e.g.,][]{howard2010}, the transit observations \citep[e.g.,][]{borucki2011}, and gravitational planetary micro-lensing \citep[e.g.,][]{cassan2012} and some in so-called habitable zones (HZs) have been already detected. 
In near future, direct detection of planetary light will play a crucial role in exploring the surface environment and meteorology of planets in habitable zones (HZ planets), and in detecting possible biosignatures there including O$_2$, O$_3$, H$_2$O absorption lines \citep[e.g.,][and references therein]{1980ASSL...83..177O,1986Natur.322..341A,1993A&A...277..309L,2002A&A...388..985S,kaltenegger2010} and the vegetation's red edge\citep[e.g.,][]{2002A&A...392..231A, woolf2002, seager2005, rodriguez2006, hamdani2006, kiang2007b, kiang2007a}. 

As a test bed for future observations of HZ planets, 
\citet{ford2001} simulated the diurnal variation of the scattered light of the Earth. 
They found 20\% (in realistically cloudy cases) or 150\% (in cloud-free cases) variations originated from the inhomogeneity of surface components and cloud cover. Their results encourage the possibility to reconstruct the landscape of exoplanets from scattered light curves.
Later on, \citet{oakley2009} suggested the inversion using the ocean glint at crescent phase and the albedo difference between ocean and land. Their methods work reasonably well when the cloud cover is lower than the actual Earth. 
\citet{cowan2009, cowan2011} performed a principal component analysis (PCA) on the multi-band photometry of the Earth obtained with EPOXI mission. They found that a dominant eigencolor of the Earth corresponds to the cloud-free land component, which allows one to draw a one-dimensional map of continental distribution along the longitude. 
\citet{fujii2010, fujii2011} decomposed of the total scattered light of the Earth into several albedo models ---ocean, snow, soil, vegetation, and clouds--- and found that the extracted time variation of them roughly recovers the actual surface inhomogeneities including the distribution of vegetation as well as continents/ocean. 

Recently, \citet[][ hereafter Paper I]{kawahara2011} proposed a new inversion method named ``spin$-$orbit tomography'' or SOT, to sketch a two-dimensional map of the exoplanet's surface using both spin rotation and orbital revolution of the planet. 
This method directly maps the reflectivity over the surface with the Tikhonov regularization, which balances between the observational noise and spatial resolution of the surface. 
This was an improvement over \citet{kawahara2010}, where the inversion was regularized by the bounded variables least square method and thereby specific albedo models were required. 
We stress that the SOT retrieves the spatially-resolved image of exoplanets from the disk-integrated light curve without directly resolving of the planet. In this sense, the SOT can be a precursor to the projects which attempt to resolve planetary image directly with very long baseline, such as Hypertelescope \citep[150km array size with 150 apertures of 3 m; e.g.][]{labeyrie1999}. 
In a hypothetical case where a mock Earth with planetary obliquity 90$^{\circ }$ is in a face-on orbit, they found that the albedo map reconstructed from a single photometric band traces the distribution of clouds. 
They also showed that cloud-free surface features could be recovered from the difference of two photometric bands since the cloud reflection spectra are roughly constant over optical wavelengths \citep[e.g. Figure 8 of][]{fujii2011}. 
The planetary obliquity was also reasonably estimated within the framework of the SOT. 

This paper aims to further develop the SOT by considering various configurations, including low-obliquity planets like the Earth in inclined orbits. 
In the case where the planetary orbit deviates from face-on, 
the phase angle of the planet is changed and the anisotropy of scattering, primarily by clouds and the atmosphere, can have non-negligible effects. 
In this paper, we take account of such effects to make the methodology applicable to more general cases. 

The organization of this paper is as follows. 
In Section \ref{s:sot}, we define the geometric configuration and describe the methodology of the SOT in detail. 
In Section \ref{s:sim}, we prepare mock light curves to be applied to the demonstration of the SOT. The simulation scheme, adopted data sets, and examples of simulated light curves are found there. 
Section \ref{s:inv} shows the results of two-dimensional mapping and obliquity measurements via the SOT, and estimates instrumental requirements for future missions. 
Case studies with different configurations are demonstrated in Section \ref{s:casestudies} with examples of a  zero-obliquity planet in more inclined orbit and a tidally-locked planet (Section \ref{ss:differentgeo}) as well as Earth models with ancient continental distribution (Section  \ref{ss:differentland}). 
Finally, Section \ref{s:sum} summarizes the results of this paper and discusses the implications of the SOT.

\section{Spin-Orbit Tomography}
\label{s:sot}

In this section we describe the methodology of the SOT, which was originally proposed in Paper I. While we described essentials of the SOT in Paper I assuming the face-on orbit, in this paper we extend our method to be applicable for an arbitrary direction of the line of sight. 
This generalization adds the orbital inclination $i$ to the set of geometrical parameters. 

\subsection{Geometry}
\label{ss:geo}

\begin{figure*}[htb]
  \includegraphics[width=\linewidth]{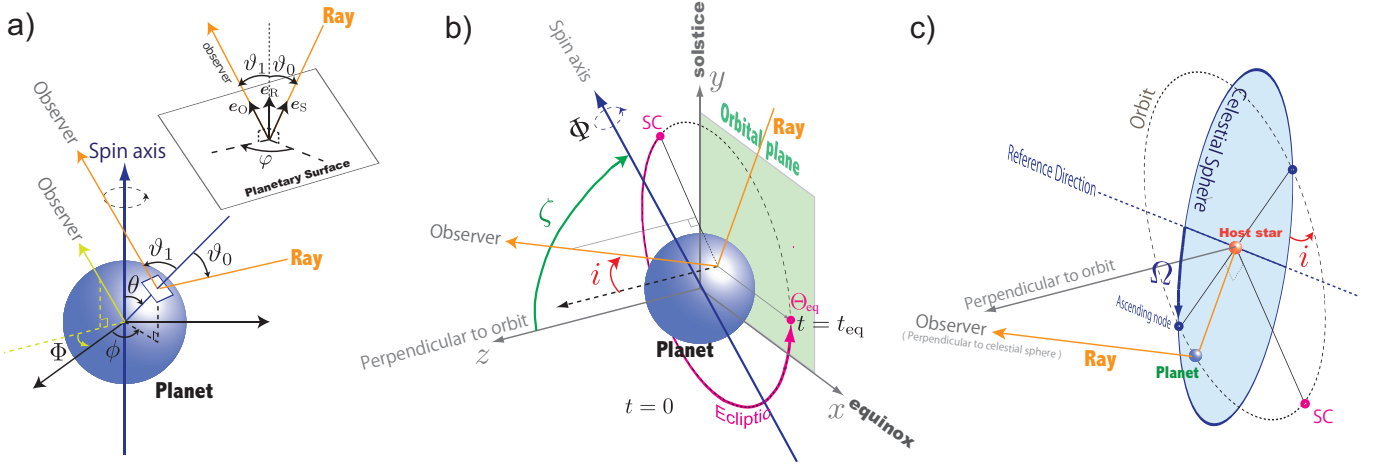}
\caption{  \label{fig:geo} Schematic configurations of the planetary system and surface. Panel (a) shows the spherical coordinate fixed on the planetary surface:  the complementary latitude, $\theta$, and the longitude, $\phi$. The spin rotation axis is indicated by $\theta=0$. We denote the spin motion by $\Phi$, the angle between ($\phi=0$, $\theta=\pi/2$) and the direction of equinox.  We define three unit vectors on the planetary surface (the normal vector, $\er$, the vector to the observer, $\elos$, and that to the host star, $\es$) which specify the arguments of the BRDF $f(\vartheta_0,\vartheta_1, \varphi ;\lambdaopt)$: the incident zenith angle, $\vartheta_0$, the zenith angle of scattering, $\vartheta_1$, and the relative azimuthal angle  between the incident and scattered light, $\varphi$.
 Panel (b) illustrates the geometrical configuration in the orbital frame. The orbit lies in the $x-y$ plane, corresponding to the directions of the equinox ($x$-axis) and the solstice ($y$-axis) of the planet. The $z$-axis is defined by the normal of the orbital plane. The obliquity $\zeta$ and the inclination $i$ are defined as the angle from the $z$-axis to the spin axis and to the line of sight, respectively. The orbital phase measured from the superior conjunction (SC) is denoted by $\Theta$. The orbital phase of equinox is indicated by $\Theta _{\rm eq}$. 
 Panel (c) displays  the geometrical configuration in the observer frame where the longitude of the ascending node $\Omega$ is indicated. }
\end{figure*}

Figure \ref{fig:geo} illustrates the geometrical configurations. We define the spherical coordinate system, $\theta$ (colatitude) and $\phi$ (longitude), fixed on the planetary surface as shown in Figure \ref{fig:geo} (a). The spin motion is specified by $\Phi (t) $, which is defined as the angle between the direction of the equinox and the origin of $\phi$. Denoting the angular velocity of the spin rotation from observation by $\omegaspin $, we can write $\Phi (t) = \omegaspin t + \Phi_{\rm offset}$. Note that $\Phi_{\rm offset} $ only changes the prime meridian and have no physical consequence. We will discuss the measurement of the spin rotation by the autocorrelation in Section \ref{ss:spin}. 
The orbital phase is described by $\Theta $, taking the superior conjunction as its origin. 
We adopt two parameters to specify the planetary spin axis: the obliquity, $\zeta $, and the orbital phase of the vernal equinox, $\Theta _\mathrm{eq}$. 
The inclination of the orbit is indicated by $i$. 

Throughout this paper, we assume that the planetary orbit is already determined, i.e., inclination $i$, the time revolution of $\Theta $, and the phase of superior conjunction are known. 
Geometrical parameterizations are summarized in Table \ref{tab:cs}. 

\begin{table}[!tbh]
\begin{center}
\caption{Coordinate System and Geometrical Configuration\label{tab:cs}}
  \begin{tabular}{cc}
   \hline\hline
Symbol & Explanation \\
\hline
   \multicolumn{2}{l}{(Coordinates)} \\
$\theta$ & Colatitude fixed on the planetary surface  \\
$\phi$ & Longitude fixed on the planetary surface \\
$\Theta$ & Orbital phase measured from superior conjunction\\
$\Phi$ & Phase of spin rotation \\ \hline
   \multicolumn{2}{l}{(Configuration of the System)} \\
 $\zeta$ & Planetary obliquity \\ 
   $i$ & Orbital inclination \\
$\Theta _{\rm eq}$ &  Orbital phase of vernal equinox \\ \hline
\end{tabular}
\end{center}
\end{table}

We introduce three unit vectors to describe the direction of the incident ray and the observer: the vector from the planetary surface to the host star, $\es$, the vector from the surface to the observer, $\elos$,  and the vector from a planetary center to the surface, $\er$ as shown in Figure \ref{fig:geo} (a). 
For mathematical simplicity, we can take an auxiliary coordinates where the $x$-axis is the direction of equinox and the $z$-axis is the orbital axis (Figure \ref{fig:geo} (b)), and the components of the three unit vectors are:
\begin{eqnarray}
  \label{eq:eh}
\es &=& (\cos (\Theta - \Theta _{\rm eq}), \sin (\Theta - \Theta _{\rm eq}), 0)^T,\\
  \label{eq:elos}
\elos &=& (\si \cos \Theta _{\rm eq}, - \si \sin \Theta _{\rm eq}, \ci)^T, \\
  \label{eq:es}
\er &=& R (\zeta) \, \erd (\phi+\Phi, \theta) \nonumber \\
&=& \left(
\begin{array}{c}
\cos{(\phi+\Phi)} \stheta \\
\czeta \sin{(\phi+\Phi)} \stheta +  \szeta  \ctheta \\
- \szeta \sin{(\phi+\Phi)} \stheta + \czeta \ctheta
\end{array} \right),
\end{eqnarray}
where  $R(\zeta)$ is the clockwise rotation matrix around $x$-axis and $\erd(\phi,\theta) = (\cphi \stheta, \, \sphi \stheta, \, \ctheta)^\trans $.

\subsection{Modeling of Scattered Light}
\label{ss:lam}

Scattering property of each patch of the planetary surface is generally expressed by the bidirectional reflectance distribution function (BRDF), $f (\vartheta_0,\vartheta_1, \varphi )$, which is a function of the incident zenith angle $\vartheta_0$, the scattering zenith angle $\vartheta_1$, and the relative azimuthal angle between the incident and scattered light $\varphi$ (if the surface have no special direction) as shown in Figure \ref{fig:geo} (a). 
The scattered intensity at each planetary pixel is expressed as
\begin{eqnarray}
\frac{d I_\lambdaopt }{d\omega } =  F_{\ast \lambdaopt} R_p^2 f_{\lambdaopt }(\vartheta _0, \vartheta _1, \varphi) \cos \vartheta_0 \cos \vartheta_1, 
\label{eq:intensity}
\end{eqnarray}
where $f_{\lambdaopt } (\vartheta _0, \vartheta _1, \varphi)$ is the BRDF at wavelength $\lambdaopt $, $F_{\ast \lambdaopt}$ is the incident flux at $\lambdaopt$, $R_p$ is a planetary radius, and $d \omega = \sin \theta d \theta d \phi$. 
Note that  $\vartheta _0$, $\vartheta _1$, and $\varphi $ are functions of all of the geometrical parameters listed in Table \ref{tab:cs}.

For the reflection model, we assume that the scattering is isotropic (Lambertian). Then the BRDF becomes independent on $\vartheta _0$ , $\vartheta _1$, or $\varphi$ and
\begin{eqnarray}
f_{\lambdaopt {\rm (LAM)}} (\vartheta _0, \vartheta _1, \varphi ) = \frac{a_\lambdaopt (\phi,\theta)}{\pi}.
\end{eqnarray}
where $a_\lambdaopt (\phi,\theta)$ is the albedo at the planetary surface at wavelength $\lambdaopt$.
Under the above assumption, Equation (\ref{eq:intensity}) reduces to
\begin{eqnarray}
I_\lambdaopt &=&  \frac{F_{\ast \lambdaopt} R_p^2 }{\pi}  \int a_\lambdaopt (\phi,\theta) \WI \WV d \omega, 
\label{eq:le}
\end{eqnarray}
where the weight functions for the illuminated and visible area are defined by $\WI \equiv \mathrm{max}\{ \cos{\vartheta_0},0 \}$ and $\WV \equiv \mathrm{max}\{ \cos{\vartheta_1} ,1 \}$, respectively \citep[see also][]{cowan2009}.

Using the three unit vectors (Equations (\ref{eq:eh})$-$(\ref{eq:es})), we obtain
\begin{eqnarray}
\label{eq:wi}
\WI(\phi,\theta;\Phi,\Thetae ) &=& \mathrm{max}\{\es \cdot \er,0 \}, \\
\label{eq:wv}
\WV(\phi,\theta;\Phi ) &=& \mathrm{max}\{ \elos \cdot \er,0 \}. 
\end{eqnarray}
The explicit forms of Equations (\ref{eq:wi}) and (\ref{eq:wv}) are given in Appendix \appweight. 

We define the weight function for the illuminated and the visible area as follows:
\begin{eqnarray}
\label{eq:weight}
W(\phi,\theta;\Phi,\Thetae) \equiv \WI(\phi,\theta;\Phi,\Thetae) \, \WV(\phi,\theta;\Phi).
\end{eqnarray}
With this weight function, Equation (\ref{eq:le}) reduces to
\begin{eqnarray}
I_\lambdaopt (\Phi, \Thetae ) =  \frac{F_{\ast \lambdaopt} R_p^2}{\pi} \int  W(\phi,\theta;\Phi, \Thetae ) a_\lambdaopt (\phi,\theta) d \omega .
\label{eq:intensity_at}
\end{eqnarray}
The integral of the weight function over the planetary surface provides the phase function of a uniform Lambert sphere \citep[e.g.,][]{russell1906},
\begin{eqnarray}
\Psi _{\rm LAM} (\alpha) &=& \int W \,d \omega \nonumber \\ 
 &=& \frac{2}{3} [\sin{\alpha} + (\pi - \alpha) \cos{\alpha}], \label{eq:phasefunc_Lam} 
\end{eqnarray}
where $\alpha $ is the phase angle, defined as 
\begin{eqnarray}
\label{eq:phaseangle}
\cos \alpha = \es \cdot \elos = \sin{i} \cos{\Theta }.
\end{eqnarray}

\begin{figure*}[!tbh]
  \begin{minipage}{0.32\linewidth}
    \begin{center}
      \includegraphics[width=\linewidth]{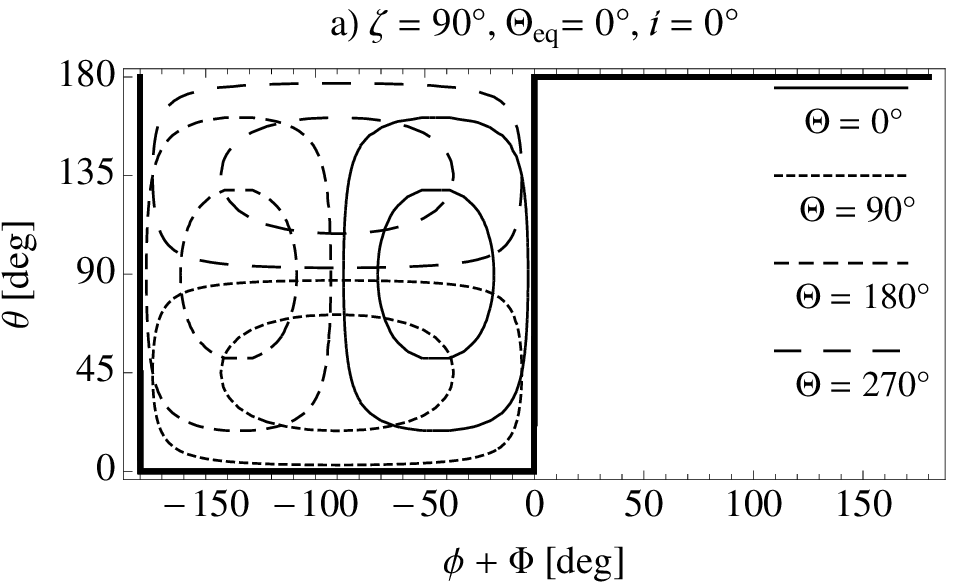}
    \end{center}
  \end{minipage}
  \begin{minipage}{0.32\linewidth}
    \begin{center}
      \includegraphics[width=\linewidth]{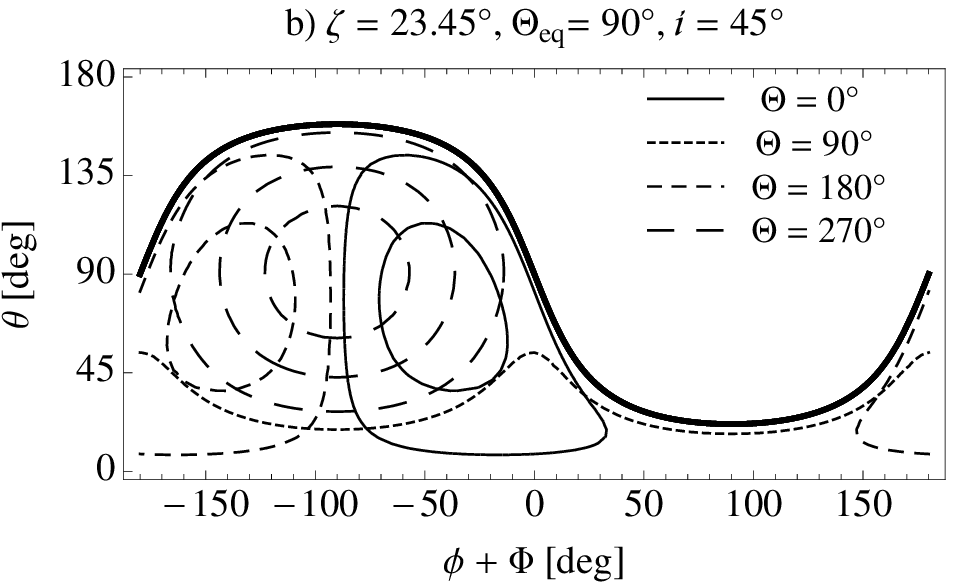}
    \end{center}
  \end{minipage}
  \begin{minipage}{0.32\linewidth}
    \begin{center}
      \includegraphics[width=\linewidth]{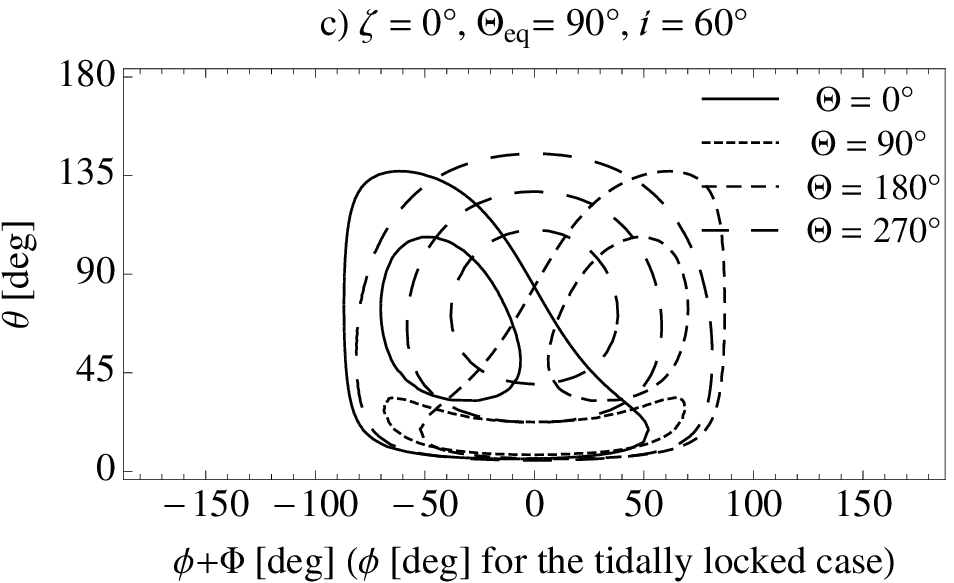}
    \end{center}
  \end{minipage}
 \caption{Examples of geometrical dependence of the weight function $W(\phi,\theta;\Phi;\Thetae)$. Annual variations are indicated by different thin lines. Solid, large-dashed, middle-dashed and small-dashed lines indicate $\Theta=0^\circ, 90^\circ, 180^\circ, $ and $270^\circ$, respectively. Inner and outer contours indicate $W(\phi,\theta; \Phi; \Theta)=0.05, 0.3$ and $0.6$. Thick lines indicate the boundary of the visible area, $\theta = \theta_{V,\pm}  (\phi)$ in Equation (\ref{eq:thetaV}). 
Panel (a) indicates an extreme case: $(\zeta=90^\circ,\Theta_\mathrm{eq}=0^\circ,i=0^\circ) $, corresponding to the geometry considered in Paper I. 
Panel (b) corresponds to the fiducial case of the Earth-twin we consider in this paper: $(\zeta=23^\circ .45,\Theta_\mathrm{eq}=270^\circ,i=45^\circ)$. 
Since the weight function solely depends on $\phi + \Phi$, in other words, longitude minus the phase of the spin rotation, we adopt $\phi + \Phi$ to the $x-$axes in panels (a) and (b) instead of showing diurnal variations explicitly.   
Panel (c) indicates the upright case  or the tidally locked case: ($\zeta = 0^{\circ },\Theta _{\rm eq}=270^{\circ },i = 60^{\circ }$). In the tidally locked case, the $x-$axis becomes
$\phi $ and does not depend on $\Phi $. \label{fig:gref}}
\end{figure*}

Figure \ref{fig:gref} displays  the behavior of the weight function in the case of $(\zeta = 90^{\circ }, \Theta _{\rm eq} = 0^{\circ }, i = 0^{\circ })$, $(\zeta = 23^{\circ }.45, \Theta _{\rm eq}= 90^{\circ }, i=45^{\circ })$, and $(0^{\circ }, 90^{\circ }, 60^{\circ })$. 
The weight function covers the region of $0 < \theta < \theta_{V,-}$ ( or $\theta_{V,+} < \theta < \pi $ )  and $-180^\circ < \phi < 180^\circ $ as the planet rotates around the host star and its spin axis, where $\theta_{V,\pm}$ is the boundary of $\WV$ described in Appendix \appweight. 

\subsection{Solving Inverse Problem \label{ss:solveinv}}

In order to linearize Equation (\ref{eq:intensity_at}), we discretize the surface to $\Mmodel$ small pixels. 
The center and the area of the $j$th pixel are denoted by $(\phi_j,\theta_j)$ and $s_j$, respectively. 
Equation (\ref{eq:intensity_at}) is then expressed as
\begin{eqnarray}
\label{eq:IFW}
I_\lambdaopt (\Phi, \Theta) &=& \sum_{j=1}^{\Mmodel} \int_{s_j} F_{\ast \lambdaopt} R_p^2 \frac{a(\phi,\theta)}{\pi}  W(\phi,\theta;\Theta ;\Phi ) d \omega . \nonumber \\
\end{eqnarray}
We assume that the pixel size is small enough to satisfy
\begin{eqnarray}
\int_{s_j} F_{\ast \lambdaopt} R_p^2 \frac{a(\phi,\theta)}{\pi}  W(\phi,\theta;\Theta, \Phi) d \omega \nonumber \\
\approx F_{\ast \lambdaopt} R_p^2 \sum _j \frac{a(\phi_j,\theta_j)}{\pi}  W(\phi_j,\theta_j;\Theta, \Phi) \Delta \omega_j, 
\label{eq:appsmall}
\end{eqnarray}
where $\Delta \omega _j$ denotes the solid angle of $j$th pixel. 

We consider $\Ndata$ bins of observational data. We define the $i$-th data point ($i=1,2,...,\Ndata$) as 
\begin{eqnarray}
d_i &=& \frac{\pi I_\lambdaopt (\Phi _i, \Theta _i)}{F_{\ast \lambdaopt }} + \error_i , 
\label{eq:dGma}
\end{eqnarray}
where $\error_i$ is noise of the $i$-th data point. 

After these discretizations, we obtain a linear discrete equation
\begin{eqnarray}
\label{eq:dGm}
\dv &=& \Dm \, \mv + \errorv, \\
 \Dm_{ij} &\equiv&  W(\phi_j,\theta_j;\Theta _i, \Phi _i) \Delta \omega_j, \\
m_j &\equiv& R_p^2 a_\lambdaopt (\phi_j,\theta_j).
\end{eqnarray}

As a pixelization scheme for the planetary surface, Hierarchical Equal Area isoLatitude Pixelization (HealPix) is adopted in this paper\footnote{http://healpix.jpl.nasa.gov/index.shtml}. 
Unless otherwise noted, the resolution parameter for HealPix $N_{\rm side}$ is set to 8 corresponding pixel numbers of which is  $\Mmodel = 12N_{\rm side}^2$ = 768 \citep{2005ApJ...622..759G}. 

Note that when the orbit is not circular one needs to caution about the incident flux $F_{\ast \lambdaopt }$ because it changes in time. 
However, the procedure of analysis after translating the observed planetary intensity into $\dv $ and the observing time into $\{\Theta , \Phi \}$ remains essentially same. 

\subsubsection{Tikhonov Regularization}
\label{sss:tikhonov}

For the time being, we assume that all geometrical parameters (i.e. not only $i $ but also $\zeta $ and $\Theta _{\rm eq}$) are known. 
To solve the linear inverse problem (Equation (\ref{eq:dGm})), we add a regularization term to the misfit function; the regularized solution is given by minimizing 
\begin{eqnarray}
Q = \sum_{i=1}^{\Ndata} \frac{|d_i - \Dm_{ij} m_j |^2}{\sigma_i^2}  + \lambdar^2 P(\mv) , 
\label{eq:qminimization}
\end{eqnarray}
where $\sigma _i ^2$ is variance of the observational error and $P(\mv)$ is the penalty function, which measures the regularity of the solution. 
The regularization parameter $\lambdar$ controls effective degree of freedom of the model. 

In this paper, we adopt the Tikhonov regularization, also known as the dumped least-square method \citep[e.g.,][]{menke1989,tarantola2005,hansen2010}. The penalty function of the Tikhonov regularization is:
\begin{eqnarray}
P_\mathrm{tik} (\mv) \equiv |\mv - \mpr|^2,  \label{eq:penalty}
\label{eq:tikhonov}
\end{eqnarray}
where $\mpr$ is the prior of the model, each component of which is fixed to the average of apparent albedo \citep[e.g.,][]{qiu2003} times $R_p^2$. 
The solution which minimizes Equation (\ref{eq:qminimization}) with Equation (\ref{eq:penalty}) is given by
\begin{eqnarray}
\label{eq:tikres}
\mestl &=& V \Sigma_\lambdar U^\trans (\tdv - \tG \, \mpr) + \mpr , \\
(\Sigma_\lambdar)_{ij} &\equiv& \frac{\eig_i}{\eig_i^2 + \lambdar^2} \delta_{ij}, 
\end{eqnarray}
where we define $\tilde{d}_i \equiv d_i/\sigma_i$ and $\tG_{ij} \equiv G_{ij}/\sigma_i$.  The orthogonal matrices $U$ and $V$ are given by the singular value decomposition of $\tG = U \Lambda V^\trans$ and $\eig_i$ is the $i$th eigenvalue of $\tG$, that is, the $i$-th component of the diagonal matrix $\Lambda$ \citep[e.g.,][]{menke1989,hansen2010}. The $\delta_{ij}$ is the Kronecker delta.

If data have enough information to reconstruct the model, $\mpr$ has little influence on $\mestl$, while if data have almost no information, $\mestl$ converges with $\mpr$. From the Bayesian viewpoint, the Tikhonov regularization is equivalent to the parameter estimation with a Gaussian-type prior with the average $\mpr$ and the covariance matrix $\lambdar^{-2} I$. 
We briefly summarize the statistics of the Tikhonov regularization in Appendix \appbayes. 

The probability that each pixel illuminates depends on the obliquity, the direction of the observer, and the latitude of the pixel. 
During a year of the planetary system, some pixels are frequently sampled and others are not. 
In order to quantify the relative influence of data on estimation of $\mestl $, we introduce the {\it integrated sensitivity vector} ${\bf S}$ \citep{zhdanov2002} defined as
\begin{equation}
S_j = \sqrt{\sum _{i=1}^{\Ndata} G_{ij}^2}.
\end{equation}
The estimated value of $m_j$ is sensitive to data if $S_j$ is large and vice versa. 
Clearly, $m_{{\rm est}, \lambdar, j} $  should be equal to $\hat m _j$ if $S_j$ is 0. 

The regularization parameter $\lambdar$ is chosen based on ``the L-curve criterion'', which is the maximum curvature point of the model norm $|\mestl - \mpr|$ versus residuals $|\tilde{\dv} - \tilde{G} \mestl|$ plot \citep{hansen2010}. 
The details of the L-curve criterion is described in Appendix \applcurve. 

\subsection{Obliquity and Equinox}
\label{ss:oblqeqnx}

So far we have described the methodology to reconstruct the albedo maps with known geometric parameters. 
In reality, the weight function $W$ and thus the design matrix $\Dm $ are functions of $\zeta $ and $\Theta _{\rm eq}$ as well: 
\begin{eqnarray}
\label{eq:dGm_zetaetimate}
\dv &=& \Dm (\zeta , \Theta _{\rm eq}) \, \mv + \errorv, \\
 \Dm_{ij} (\zeta , \Theta _{\rm eq}) &\equiv&  W(\phi_j,\theta_j;\Theta _i, \Phi _i; \zeta , \Theta _{\rm eq}) \Delta \omega_j. 
\end{eqnarray}
While direct imaging will be able to estimate $i$ and the superior conjunction, two parameters which specifies spin axis (obliquity, $\zeta $, and orbital phase of equinox, $\Theta _{\rm eq}$) are difficult to estimate by any other known techniques. 
In face-on cases, Paper I showed that these two parameters can be simultaneously estimated via the SOT itself by taking them as free parameters and minimizing $Q$%
\footnote{%
In the case of face-on orbit discussed in Paper I, $\Theta $ was measured from the vernal equinox instead of the point of superior conjunction because the latter cannot be defined, and accordingly the measurement of $\Theta _S$, the phase of the start of observation relative to the equinox, was to be estimated rather than $\Theta _{\rm eq}$. %
}. 
In the same manner, we search for the best-fit $\zeta$ and $\Theta_\mathrm{eq}$ by the Nelder-Mead method with varying $\lambda $. At each trial value of $\zeta $ and $\Theta_\mathrm{eq}$, the best-fit model vectors $m_j$ are derived by Equation (\ref{eq:tikres}). 
The set of the best-fit $\zeta$, $\Theta_\mathrm{eq}$ and $m_j$ for different $\lambda $ allows us to draw the L-curves, and hence we can finally  determine the appropriate $\lambda $ by the L-curve criterion.

\section{Simulation of scattered light curves of an Earth-twin}
\label{s:sim}

In this section, we describe our Earth model to be input to the SOT as well as the resultant light curves. 

So far, various authors have carried out simulations of scattered light curves of an Earth-twin \citep{ford2001, tinetti2006a, tinetti2006b, rodriguez2006, palle2008, oakley2009, arnold2009, doughty2010, fujii2010,  fujii2011}.  
In this paper, we follow the simulation scheme described by \citet{fujii2011}; we (1) pixelize the surface of the planet into $2^{\circ }.0 \times 2^{\circ }.0$, (2) assign the parameters of the scattering properties to each patch according to the data obtained with the MODIS (MODerate resolution Imaging Spectroradiometer) onboard the Earth Observing Satellites {\it Terra}/{\it Aqua}, (3) calculate the radiative transfer in each patch by {\it rstar6b} \citep{nakajima1988, nakajima1983}, and (4) integrate the emergent light from each patch over the illuminated and visible portion. 

Since we do not have a specific instrumental design in mind, we simply assume three $0.1\mu $m width bands: 0.4$-$0.5$\mu $m ({Blue), 0.6$-$0.7$\mu $m (Orange), and 0.8$-$0.9$\mu $m (NIR). 
Since it is fairly time consuming to sum up the radiance over planetary surface and to integrate fine spectrum over photometric bands, we create a look-up table with relevant parameters in the same way as \citet{fujii2011} and estimate the radiance at a given pixel by linearly interpolating it. 

In Section \ref{ss:input}, we describe the adopted data sets and our assumptions in more detail. 
Examples of simulated light curves are shown in Section \ref{ss:lc}. 

\subsection{Assumptions and Input Data}
\label{ss:input}

The {\it rstar6b} calculates the radiative transfer through the atmospheric layers given the properties of atmosphere, ground albedo,  
the direction of the host star, and that of the observer. 
The vertical profiles of pressure, temperature, and molecule composition in the atmosphere are assumed to be uniform for simplicity; we adopt the US standard model.\footnote{U.S. Standard Atmosphere, 1976, U.S. Government Printing Office, Washington, DC, 1976}
The thermal profile affects little on the resultant spectra at the wavelengths where the thermal radiation is not dominant. 

Since cloud cover significantly changes the reflection spectra, we insert a cloud layer according to the actual cloud data of each pixel. 
In our model, cloud cover is specified by two parameters: cloud cover fraction $f_{\rm cld}$ and cloud optical thickness $\tau _{\rm cld}$. The daily global map for these two parameters is taken from the Terra/MODIS Atmosphere Level 3 Product\footnote{http://ladsweb.nascom.nasa.gov/}. Specifically, we adopt the data of 2008. 
Other parameters for cloud optical properties are fixed to the value adopted by \citet{fujii2011}. 
In particular, we assume that clouds are composed of pure liquid water and that the cloud layer is at 3.5$-$6.5 km. 
The altitude of the cloud layer affects the continuum level little \citep[e.g. Figure 8 of][]{fujii2011}. 

The boundary condition of the radiative transfer is determined by the BRDF of the surface. 
The MODIS project adopts the {\it Ross-Li} model to specify the BRDF of land surface \citep[e.g.,][]{lucht2000} and offers the three coefficients in the Ross-Li model (coefficients of isotropic, geometric, and volume terms). 
We, however, use the isotropic term only in our simulation to match the specification of {\it rstar6b}. 
We obtain the value of the coefficient of the isotropic term at each pixel from ``snow-free gap-filled MODIS BRDF Model Parameters''. These data sets are a spatially and temporally averaged product derived from the 0.05 resolution BRDF/albedo data (v004 MCD43C1)\footnote{%
Available at http://modis.gsfc.nasa.gov/%
}. 
The original data sets are merged into the adopted $2^{\circ }.0\times 2^{\circ }.0$ resolution by averaging. 

In order to incorporate the effect of seasonal variation of snow cover, we adopt the monthly snow cover map offered by MODIS project.\footnote{%
http://modis-snow-ice.gsfc.nasa.gov/%
} 
We replace surface albedo by the reflection spectrum of fine snow for pixels with snow cover fraction larger than 50\%. 

In short, our model in this paper incorporates the effect of time variation of the cloud cover fraction, cloud optical thickness, surface BRDF, and snow cover. 

\subsection{Light Curves}
\label{ss:lc}

\begin{figure}[!h]
  \centerline{\includegraphics[width=80mm]{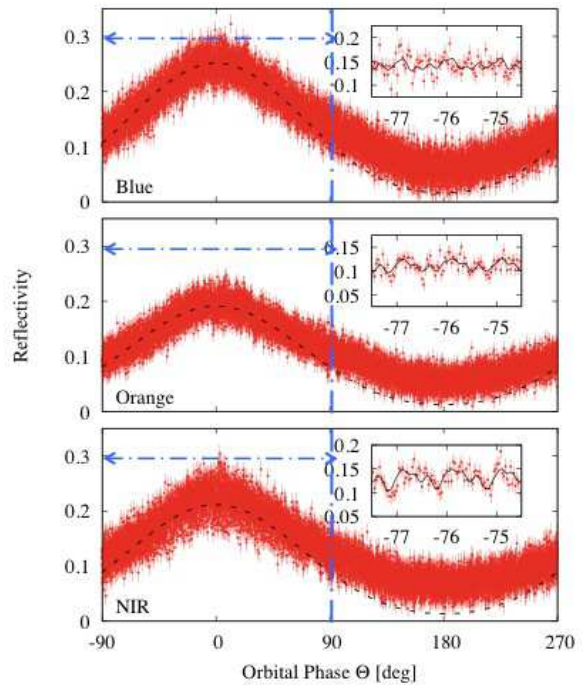}}
\caption{Yearly light curves of an Earth-twin in the case of $(\zeta = 23^{\circ }.45$, $\Theta _{\rm eq }=90^{\circ}$, $i=45^{\circ })$. 
Signal-to-noise ratio (S/N) $20/\sqrt{6}$ is imposed so that S/N after folding the light curves for 6 days becomes 20).  
Top: light curve of Blue (0.4$-$0.5$\mu $m) band and phase curves of a Lambert sphere with albedo 0.33. Middle: light curve of Orange (0.6$-$0.7$\mu $m) band and phase curves of a Lambert sphere with albedo 0.25. Bottom: light curve of {\it NIR} (0.8$-$0.9$\mu $m) band and phase curves of a Lambert sphere with albedo 0.28.
The inserted panels enlarge the diurnal variations at $\Theta \sim 15 ^{\circ }$ and show theoretical light curves by solid lines for reference. In the analysis in Section \ref{s:inv}, data within the region bracketed by dot-dashed arrows are used. }
\label{fig:lc}
\end{figure}

\begin{figure}[!h]
  \centerline{\includegraphics[width=80mm]{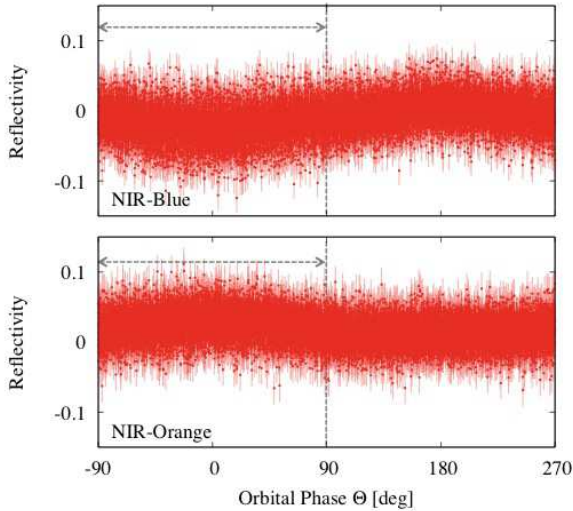}}
\caption{Top: yearly variations of the difference between the reflectivity in NIR (0.8$-$0.9$\mu $m) and that in Blue (0.4$-$0.5$\mu $m), both of which are shown in Figure \ref{fig:lc}. The geometric parameter set is $(\zeta = 23^{\circ }.45$, $\Theta _{\rm eq}=90^{\circ}$, $i=45^{\circ })$. Bottom: same as the top panel but the difference between the reflectivity in NIR (0.8-0.9$\mu $m) and that in Orange (0.6-0.7$\mu $m). }
\label{fig:lc_diff}
\end{figure}

Figure \ref{fig:lc} displays the resultant light curves of the Earth in three photometric bands with signal-to-noise ratio (S/N) = $20/\sqrt{6}$ in the case of $(\zeta = 23^{\circ }.45, \Theta _{\rm eq} = 90^{\circ }, i=45^{\circ })$.  
The inserted panels show the diurnal variations and theoretical (noiseless) light curves in black dotted lines. 
The variation is due to the spin rotation and reflects the longitudinal inhomogeneity. The amplitude of variation ($\sim 20$\%) is consistent with former works \citep{ford2001, oakley2009}. 

The yearly variation is due to the change in the illuminated and visible area (Section \ref{s:sot}), i.e., waxing and waning of the planet. 
In Figure \ref{fig:lc}, yearly variations of a hypothetical Lambert sphere are also plotted by dashed lines for reference. 
Simulated light curves substantially exceed the Lambert phase curves at phase $\Theta = 180^{\circ }$, corresponding to the maximum phase angle. 
The excess at this phase is likely due to the enhanced forward scattering by clouds and atmosphere \citep[e.g.,][]{robinson2010}. 
This deviation from isotropic scattering possibly causes a systematic bias on the reconstructed map. 
Unless the change of phase angle is negligible, one must consider this anisotropic effect. 
In this paper, we avoid this problem in a simple way: we do not use the data at crescent phase since the fractional deviation from Lambert sphere is quite significant at this phase (see Section \ref{ss:Earth_map} below for further discussion). 

Figure \ref{fig:lc_diff} shows the differential light curves between two photometric bands (NIR$-$Blue, NIR$-$Orange) for later use.

\section{Global mapping of an Earth-twin}
\label{s:inv}

In this section, we apply the SOT to the simulated light curves. 
In Section \ref{ss:spin}, we begin with pre-processing of data to be relevant to the SOT by measuring of the spin rotation period. 
A two-dimensional mapping is discussed in Section \ref{ss:Earth_map}. 
The obliquity measurement is studied in Section \ref{ss:Earth_geoparam}. 
The desired aperture size for reasonable mapping via the SOT is estimated in Section \ref{ss:noise}. 

\subsection{Pre-processing}
\label{ss:spin}

\begin{figure}[!h]
  \begin{center}
      \includegraphics[width=70mm]{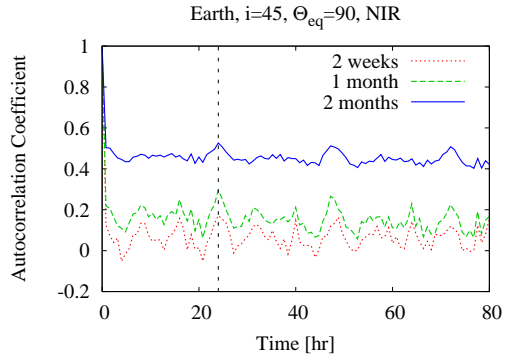}
  \end{center}
  \caption{Autocorrelation coefficients computed from the first 14 day  (red solid), 30 day (green long-dashed), and 60 day (blue short-dashed) mock observations in {\it NIR} in the case of $(\zeta = 23^{\circ }.4, \Theta _{\rm eq} = 90^{\circ }, i=45^{\circ })$ (Figure \ref{fig:lc}). }
\label{fig:autocorr}
\end{figure}

First of all, we consider translation of the observation time to phase angles $\{\Theta , \Phi \}$. 
Orbital phase $\Theta $ is known once the planetary orbit is determined, which is one of the assumptions in this paper. 
On the other hand, spin phase $\Phi $ needs to be known by measuring spin period $\omega _{\rm spin}$. 
Therefore, we perform auto-correlation analysis on our mock data as suggested by \citet{palle2008}. 

Figure \ref{fig:autocorr} displays the auto-correlation coefficients from 0.8 to 0.9 $\mu $m  data obtained in the first 7 day, 30 day, 60 day mock observations. 
We find that the spin rotation period of the Earth, 24 hr, is safely measured from any of single-band observations for $\sim $ 1 month, which enable us to assign $\Phi $. 
Note that the autocorrelation coefficient at $t \not = 24$ (hr) is not approaching 0 as the number of days increases because of the phase variation. When $i=0$, the phase angle does not change and the autocorrelation is conversing to 0. 

Furthermore, we stack the light curves for every 6 days to increase the signal-to-noise ratio per data point. 
After all we are left with $30 \times (366/6) = 1830$ data sets of $\{\Theta ,\Phi, I_i, \sigma _i \}$ for each band ($N$=1830) with S/N=20. 

\subsection{Two-dimensional Mapping}
\label{ss:Earth_map}

\begin{figure*}[!h]
    \begin{center}
      \includegraphics[width=\hsize]{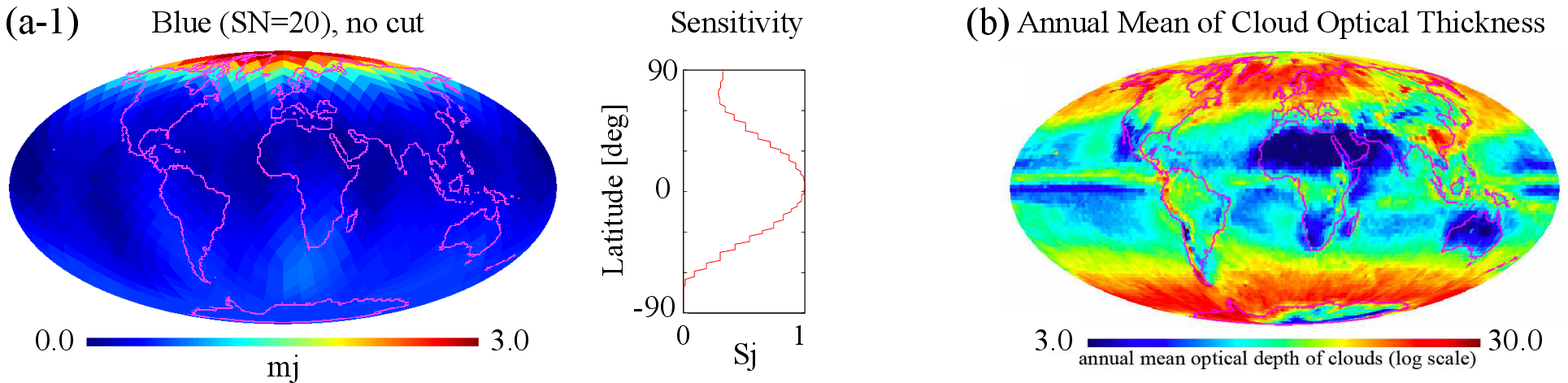}
      \includegraphics[width=\hsize]{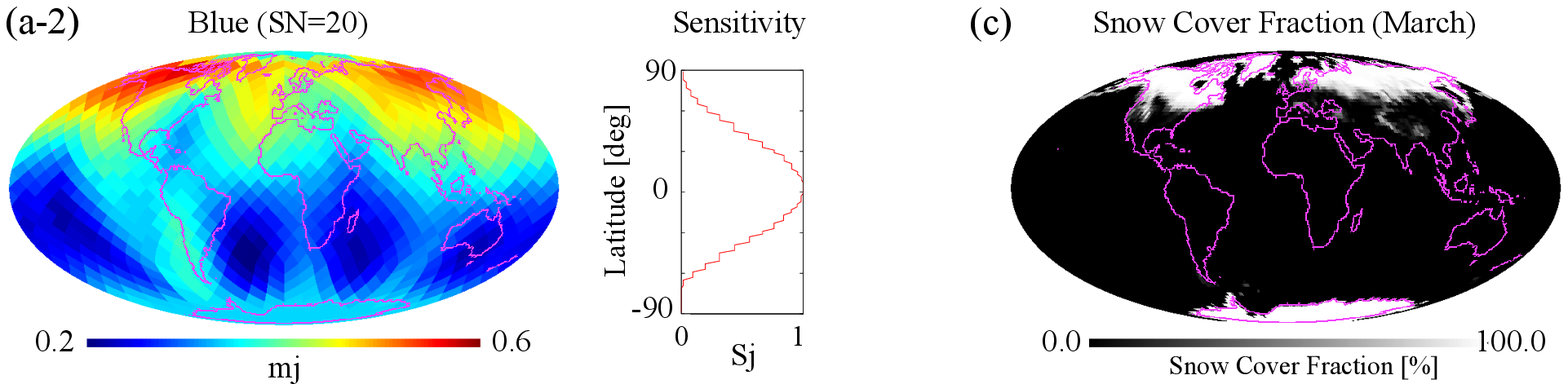}
      \includegraphics[width=\hsize]{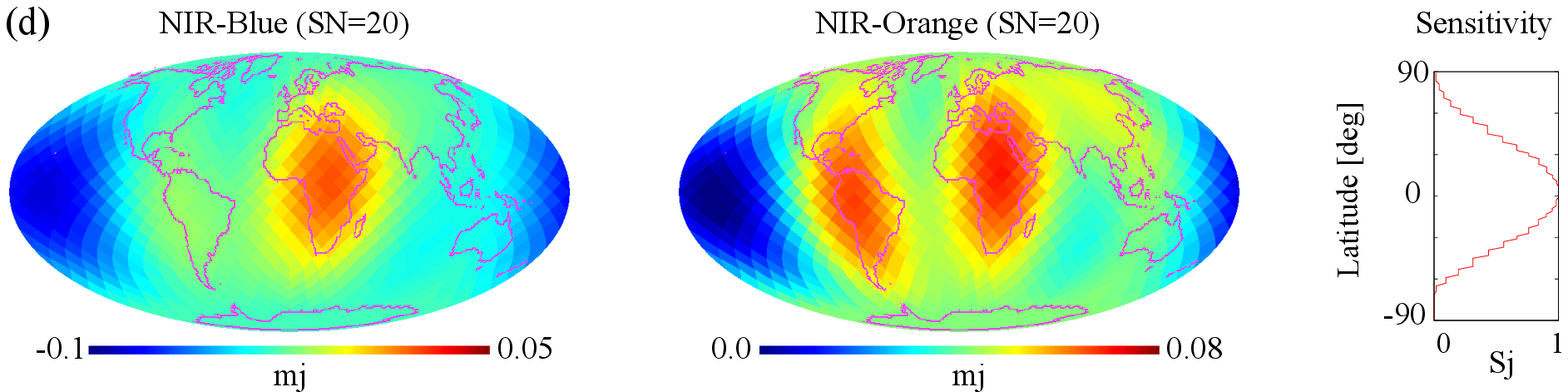}
      \includegraphics[width=\hsize]{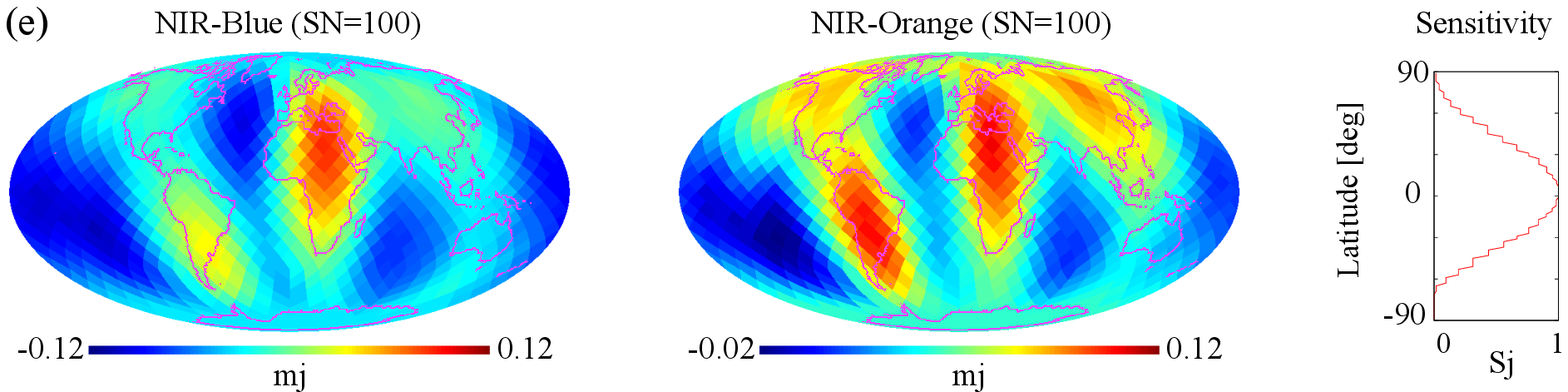}
    \end{center}
  \caption{(a-1) Two-dimensional mapping from one-year observation in Blue band with S/N=20 in the case of $(\zeta = 23^{\circ }.4, \Theta _{\rm eq}=90^{\circ }, i=45^{\circ })$ without consideration of bias caused by anisotropic scattering, i.e., data at all phases are used in the analysis. 
(a-2) Same as the top panel but data at $90^{\circ } \le \Theta \le 270^{\circ }$ are not used in the analysis since the effect of forward scattering of clouds is not negligible. 
(b) Annual mean of cloud optical thickness. 
(c) Cloud cover fraction in 2008 March. 
(d) Two-dimensional mapping from the difference between two photometric bands with S/N=20.
(e) Two-dimensional mapping from the difference between two photometric bands with S/N=100.
Plots of integrated sensitivity $S_j$ as a function of latitude are attached for respective cases. Note that $S_j$ does not depend on longitude because spin rate is much faster than orbital motion. }
\label{fig:map_Earth_inc30_tmu0_sn20}
\end{figure*}

We now adopt the SOT to mock light curves of the Earth. 
First, we naively used one-year light curves without considering the anisotropic effect. 
The resultant map displayed in panel (a-1) in Figure \ref{fig:map_Earth_inc30_tmu0_sn20} exhibits a distinct intense region at the North pole and hardly traces any real features of the Earth. 
This is due to the fact that the reflectivity at large phase angle ($\Theta \sim 180^{\circ }$ see also Equation (\ref{eq:phaseangle}])) is much larger than expected from Lambertian phase curves (see Figure \ref{fig:lc}). In reality, the increase in reflectivity at $\Theta \sim 180^{\circ }$ originates from the forward scattering of clouds and atmosphere as mentioned above. This result implies that we cannot ignore the bias from the anisotropy of scattering in the case of inclined orbit. In contrast, the anisotropy mostly leads to the offset for the face-on case ($i=0$) discussed in Paper I since the phase angle is constant. 

In order to reduce this bias, we exclude the data with large phase angle $\alpha$. 
In this paper we use the data that satisfies the condition $\alpha \le 90^{\circ }$ where the deviation from the phase curve of a Lambert sphere is significant \citep[see also Figure 2 of][]{robinson2010}.  
This condition corresponds to $-90 ^{\circ } \le \Theta \le 90^{\circ }$ (the bracketed region in Figure \ref{fig:lc}) based on Equation  (\ref{eq:phaseangle}).

Panel (a-2) in Figure \ref{fig:map_Earth_inc30_tmu0_sn20} shows the albedo map reconstructed from the Blue light curve at $\alpha \le 90^\circ$. The reconstructed map exhibits the highly reflecting zone at high latitude in the north hemisphere. 
The high-albedo region at $\theta \sim 60^{\circ }$ can be interpreted as a pattern of clouds and snow, which have high reflectivity in the Blue  band. 
We can compare the reconstructed map with an annual mean distribution of cloud optical thickness (panel (b)) and snow cover fraction (panel (c)). 
The intensive region in the south hemisphere is not seen in the resultant map. 
This is a natural consequence because the region near the south pole is insensitive to the data for this specific geometry ($\zeta = 23^{\circ }.4, \Theta _{\rm eq} = 90^{\circ }, i=45^{\circ }$). 
The  {\it integrated sensitivity}, $S_j$, to data is plotted as a function of latitude at the right of each map. 
The interpretation of the reconstructed albedo map should be accompanied with the sensitivity plot. 

Surface inhomogeneity is expected to be uncovered by mapping the difference between two photometric bands since the reflection spectra of clouds are roughly constant against wavelengths from visible to near infrared bands (see Figure 3 in Paper I). 
In this case, we substitute ${\dv}_{A} - {\dv}_{B}$ and ${\mv}_{A} - {\mv}_{B}$ for $\dv$ and $\mv$ in Equations (\ref{eq:dGm})$-$(\ref{eq:tikres}) where subscripts $A$ and $B$ indicate different bands.
We again employ the criterion $\alpha \le 90^{\circ }$ for usable data because the contribution of clouds is not completely compensated by taking difference of two bands. 
The reconstructed maps from the difference between two photometric bands with S/N=20 (Figure \ref{fig:lc_diff}) are displayed in panel (d) in Figure \ref{fig:map_Earth_inc30_tmu0_sn20}. 
The map for NIR$-$Blue (left panel) can be interpreted as a proxy of the land distribution, since most of the land components have higher albedo than ocean at the near-infrared range (Figure 3 of Paper I). 
Indeed one can see the bright regions corresponding to the Africa continent and other continents (South America, North America, and Eurasia).  On the other hand, the map from NIR$-$Orange (middle panel) is sensitive to the vegetated area as well because of the vegetation's red edge. The bright regions in the NIR$-$Orange trace the vegetated area of Africa, South America, and Eurasia well. 

We emphasize the fact that the SOT maps the albedo contrast over planetary surface and the map retrieved by the SOT is independent of any assumptions of surface components. 
The interpretation of the albedo contrast is in our hands and depends on our knowledge about reflection spectra of materials. Hence the degeneracy among components with similar reflection spectra is inevitable. For instance, distinguishing clouds from snow is difficult from the bands we considered because they both can have high and flat reflection spectra in the optical range.

Panel (e) displays the same results as panel (d) except that imposed S/N per data point is 100. 
The maps with S/N$=100$ data have higher spatial resolution than those with S/N$=20$ data and the continental distribution are clearly recovered. 

In short summary, the SOT can also be applied to Earth-twins in inclined orbits with observations at the phases larger than $90^{\circ }$. These phases are also favorable from the view point of observation because the intensity of planet is larger at smaller phase angle.

\subsection{Measurements of Obliquity and Equinox}
\label{ss:Earth_geoparam}

\begin{figure*}[!bth]
  \begin{minipage}{0.32\linewidth}
    \begin{center}
      \includegraphics[width=\hsize]{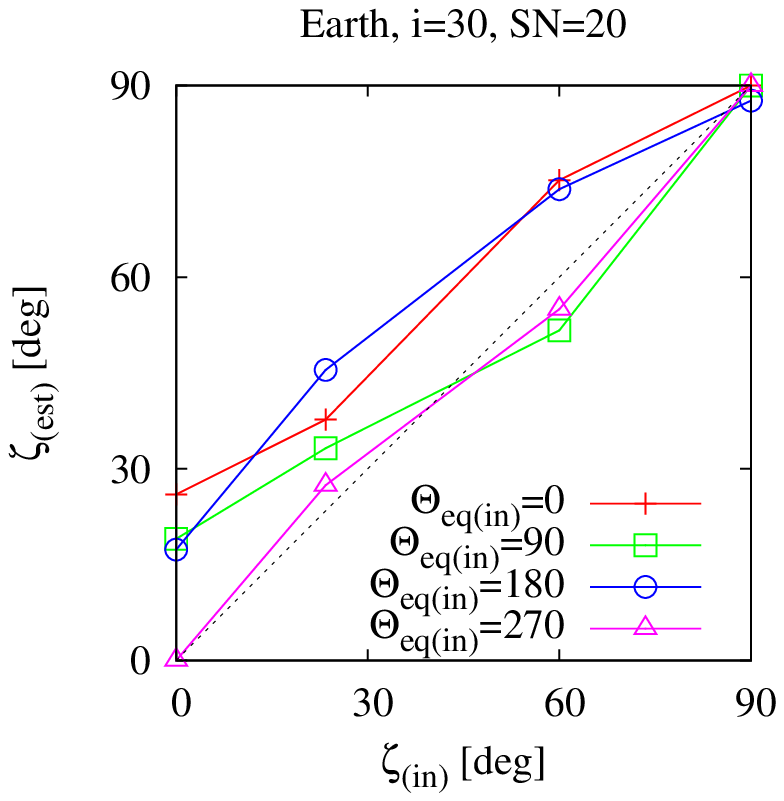}
    \end{center}
  \end{minipage}
  \begin{minipage}{0.32\linewidth}
    \begin{center}
      \includegraphics[width=\hsize]{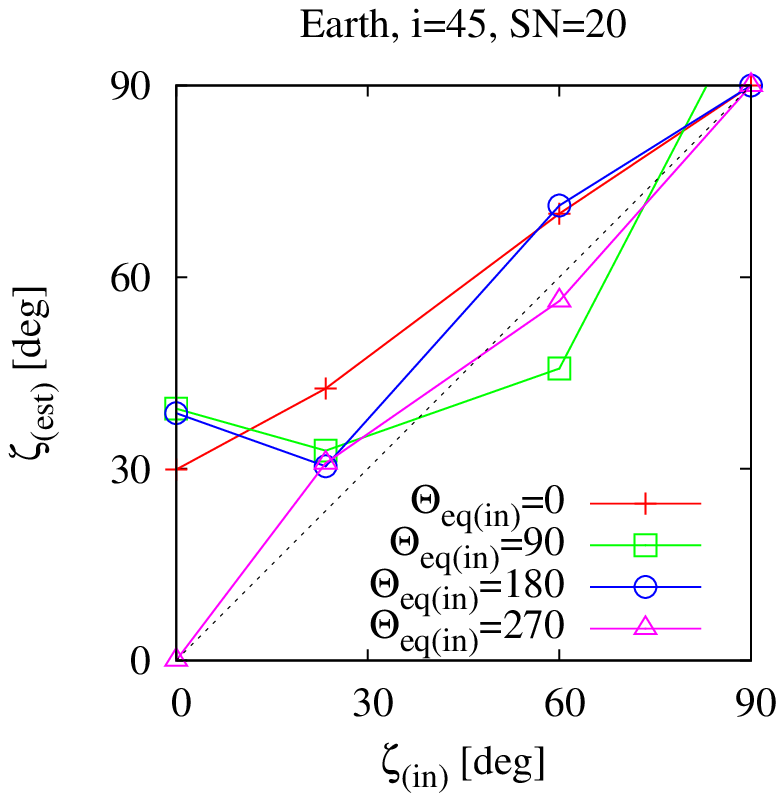}
    \end{center}
  \end{minipage}
  \begin{minipage}{0.32\linewidth}
    \begin{center}
      \includegraphics[width=\hsize]{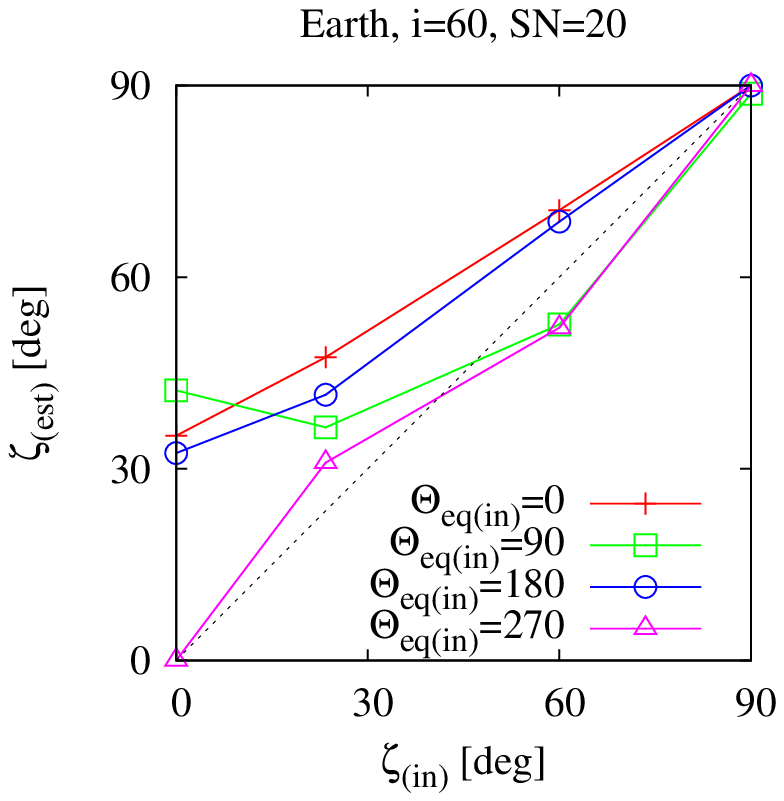}
    \end{center}
  \end{minipage}
  \begin{minipage}{0.32\linewidth}
    \begin{center}
      \includegraphics[width=\hsize]{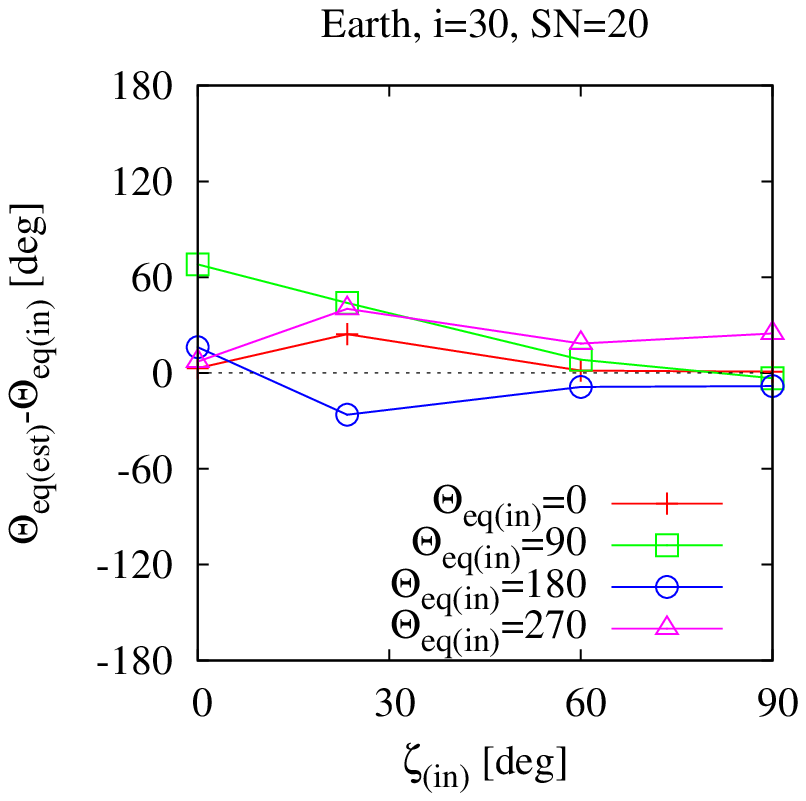}
    \end{center}
  \end{minipage}
  \begin{minipage}{0.32\linewidth}
    \begin{center}
      \includegraphics[width=\hsize]{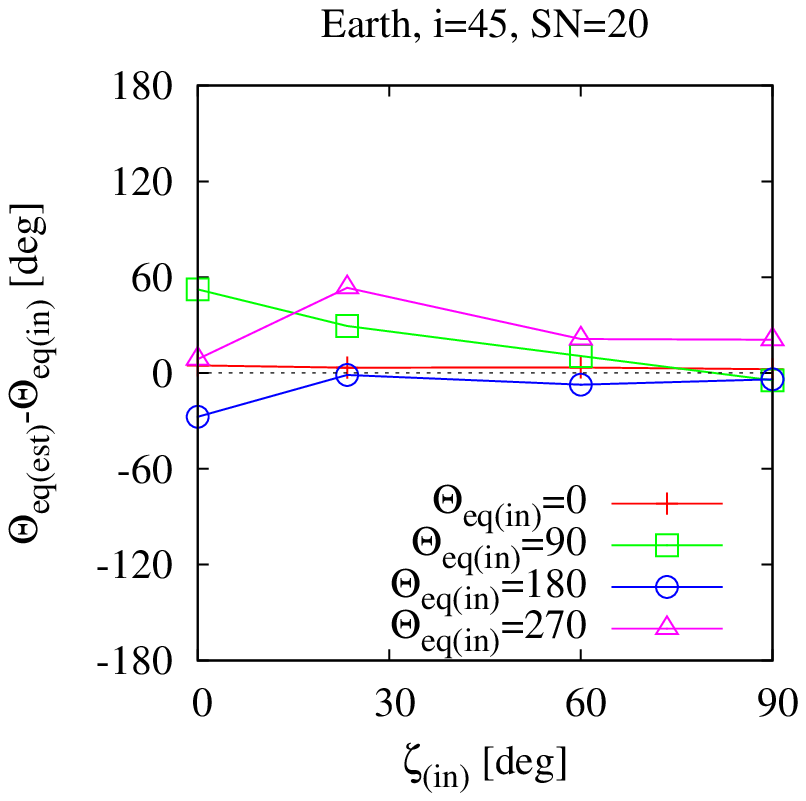}
    \end{center}
  \end{minipage}
  \begin{minipage}{0.32\linewidth}
    \begin{center}
      \includegraphics[width=\hsize]{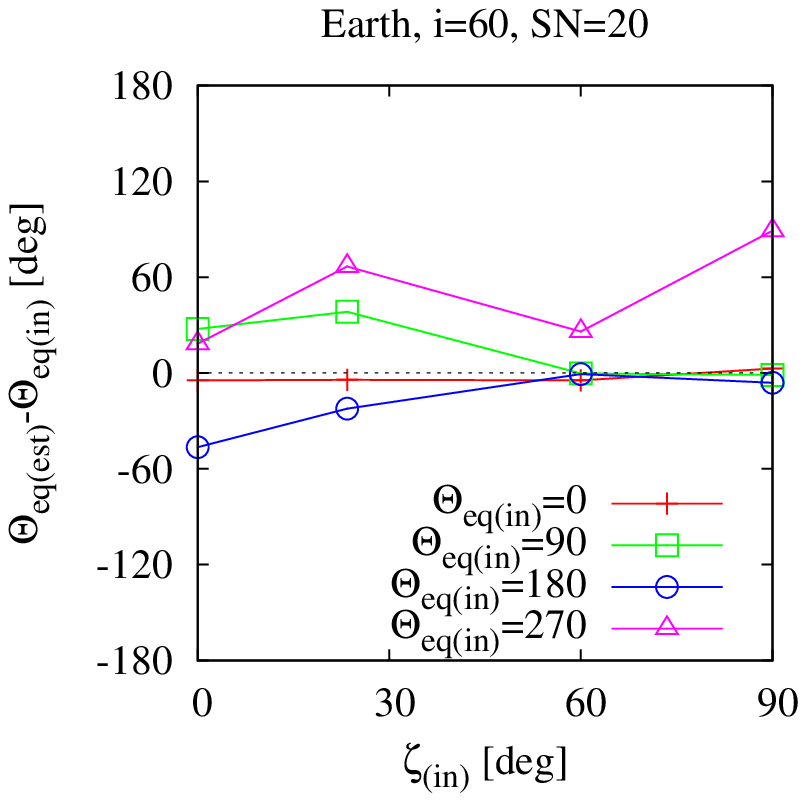}
    \end{center}
  \end{minipage}
  \caption{Estimated value of obliquity, $\zeta _{\rm (est)}$ (top row), and simultaneously estimated value of equinox phase, $\Theta _{\rm eq(est)}$ (bottom row), as a function of input value of obliquity, $\zeta _{\rm (in)}$. Different colors show the result with varying input value of equinox phase, $\Theta _{\rm eq}$. 
Three inclinations, $i=30^{\circ }$ (left column), $i=45^{\circ }$ (center column), and $i=60^{\circ }$ (right column), are studied. Signal-to-noise ratio is 20. 
}
\label{fig:zetaestimate}
\end{figure*}

In this section, we test the obliquity measurement introduced in Section \ref{ss:oblqeqnx}. 
Unlike Section \ref{ss:Earth_map}, we regard not only $m_j$ but also $\zeta $ and $\Theta _{\rm eq}$ as fitting parameters (but the orbital inclination $i $ remains to be fixed). 
We create mock light curves of Earth-twins with varying $\zeta $, $i$, $\Theta _{\rm eq}$ and perform the SOT on them respectively.  
The same data sets for cloud cover, surface reflectivity, and snow cover are adopted as input in the simulations though the planetary climates may be admittedly inconsistent with obliquity when $\zeta \not = 23^{\circ }.4$. 

Figure \ref{fig:zetaestimate} shows the measurement of $\zeta $ and $\Theta _{\rm eq}$ from the NIR$-$Blue mapping as a function of input value of $\zeta $. 
As can be seen, the estimated values are well correlated to the input values. 
For the highly oblique planets ($\zeta \sim 90^{\circ }$), the estimation works remarkably well, consistent with the face-on cases discussed in Paper I. 
Depending on the geometry of observation, there are systematic biases in some cases. 
Identifying the exact cause of the systematics is complex and we postpone the problem of precise measurement of obliquity until a future paper. 
Nevertheless, the fact that the SOT can roughly measure the obliquity and equinox is remarkable since the obliquity measurement is significantly difficult with other method. 

We also tried the same estimation with single-band light mapping, but the result is much less reliable. 
Not only latitudinal but also longitudinal inhomogeneity just like the land distribution of the Earth appears to be essential for measuring the planetary spin axis.

\subsection{Prospects with Future Instruments}
\label{ss:noise}

\begin{table*}[!bth]
\begin{center}
\caption{Observation Parameters for Noise Estimation\label{tab:noise}}
  \begin{tabular}{cccc}
   \hline\hline
Symbol	&	Quantity & Value & Unit	\\ \hline 
$t_{\rm exp}$	&	Exposure time & 24/30$\times$3600 &	s  \\ 
$n$	&	Folded days  & 6 &		days \\ 
$\Psi $	&	Sharpness	&	0.08	&	\\
$h$	&	End-to-end efficiency &	0.5	&  \\
$\upsilon$	&	Dark rate 	& 0.001	&	counts s$^{-1}$	\\
$\kappa$		&	Read noise 	& 2 		&	$\sqrt{{\rm counts}}$ read$^{-1}$	\\
QE	&	Quantum efficiency	 & 0.91	&	\\
$\Omega _z$	&	Zodiacal light in magnitude &		23	&	mag arcsec$^{-2}$	\\
$Z$	&	Zodiacal light  &	2	&	\\
\hline
\end{tabular}
\end{center}
\end{table*}

\begin{figure}[!h]
    \begin{center}
      \includegraphics[width=\hsize]{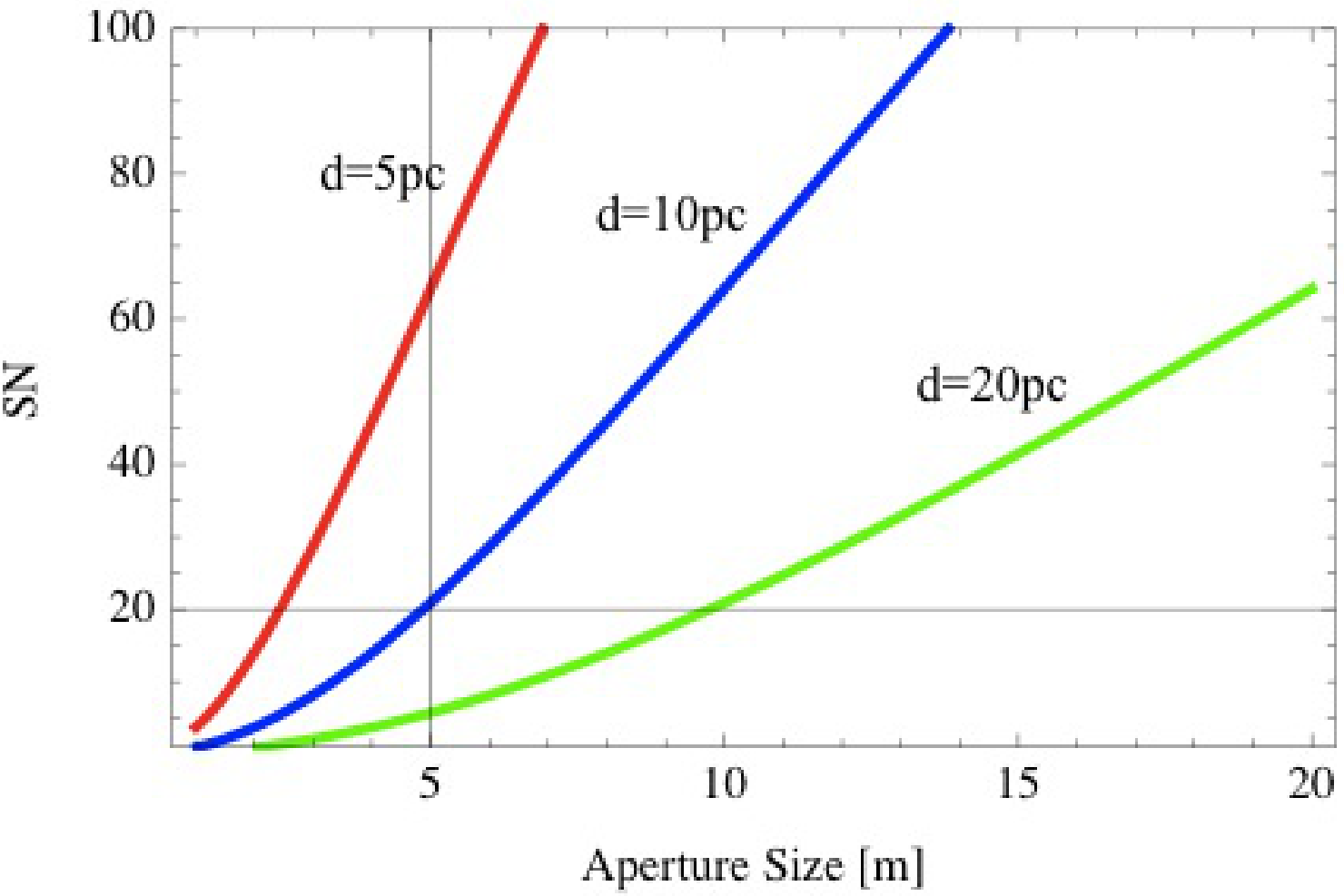}
    \end{center}
    \caption{Signal-to-noise ratio as a function of aperture size on the assumption of parameters listed in Table \ref{tab:noise}.}
\label{fig:sn-ap}
\end{figure}

Since the development for direct imaging observation is currently ongoing and not yet put into practice, we estimate the required aperture size of space telescopes for the SOT simply as follows. 
We consider readout noise, dark noise, and noise from exozodiacal light in addition to the photon noise of planetary light itself, and calculate S/N as a function of aperture size with Equations (21)$-$(24) of \citet{kawahara2010}. The leakage of the light from the host star might be a dominant noise source for some instruments but depends on the situation. 
Considering that the occulter concept can safely neglect the leakage of the host star, we do not include it in our S/N estimation. 
The input values for the parameters to specify the instrumental property are summarized in Table \ref{tab:noise}. 

Based on these assumptions, S/N of 20 with an Earth-sized planet at the distance of 10 pc will require $\sim $5 m aperture, and that of 100 will be equivalent to $\sim $ 15 m aperture size (Figure \ref{fig:sn-ap}). 
This is significantly smaller than the scale proposed for future missions aiming to obtain spatially-resolved image of exoplanets in a direct manner \citep[e.g. hypertelescope;][]{labeyrie1999}. This is the outcome of the SOT which does not resolve the image of the planet instantaneously but instead solves the inverse problem of annual light curves.

\section{Case Studies}
\label{s:casestudies}

In this section, we apply the SOT to different configurations. We consider different geometries with different combinations of orbital inclination and obliquity (Section \ref{ss:differentgeo}), and different continental distribution based on ancient Earth maps (Section \ref{ss:differentland}).

\subsection{Different Geometry}
\label{ss:differentgeo}

\subsubsection{Upright Earth-Twin}
\label{sss:upright}

\begin{figure}[!tbh]
  \begin{center}
      \includegraphics[width=\hsize]{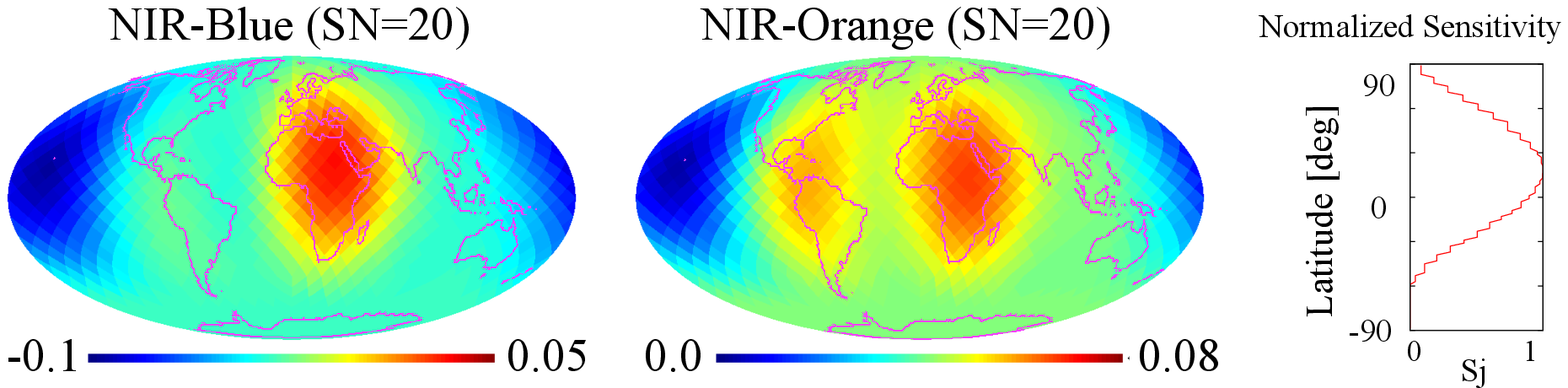}
    \end{center}
    \caption{Two-dimensional mapping from NIR$-$Blue (left) and NIR$-$Orange (right) light curves in the case of $\{\zeta , \Theta _{\rm eq}, i\}=\{0^{\circ }, 90^{\circ }, 60^{\circ }\}$. 
Data at $\alpha \le 90^{\circ }$ are used in the analysis. }
\label{fig:tikteacup}
\end{figure}

For the face-on case examined in Paper I, the two-dimensional image can be retrieved for non-zero obliquity. For inclined ($i \not = 0$) cases, however, even non-oblique planets (i.e., $\zeta  = 0$) can provide two-dimensional information of the planetary surface in principle because the visible region on the surface changes according to the orbital revolution as shown in Figure \ref{fig:gref}(c). 
To demonstrate such cases, we simulate light curves of a mock Earth with zero obliquity with orbital inclination $i = 60^{\circ }$ (Fig. \ref{fig:gref}(c)). 
In this case, we may ignore the seasonal variation of polar snow covers and surface BRDFs (we adopt the BRDFs of March all year around) but the same input data of clouds are used.

Figure \ref{fig:tikteacup} displays the resultant maps from NIR$-$Blue and NIR$-$Orange for $i=60^{\circ }$ and $\Theta _{\rm eq} = 90^{\circ }$. 
They shows as good correspondence to the real surface inhomogeneity although the resolution is slightly worse than the case discussed in Section \ref{s:inv}.

\subsubsection{Tidally Locked Earth-Twin}
\label{sss:tidal}

\begin{figure}[!tbh]
  \begin{minipage}{0.484\linewidth}
    \begin{center}
      \includegraphics[width=\linewidth]{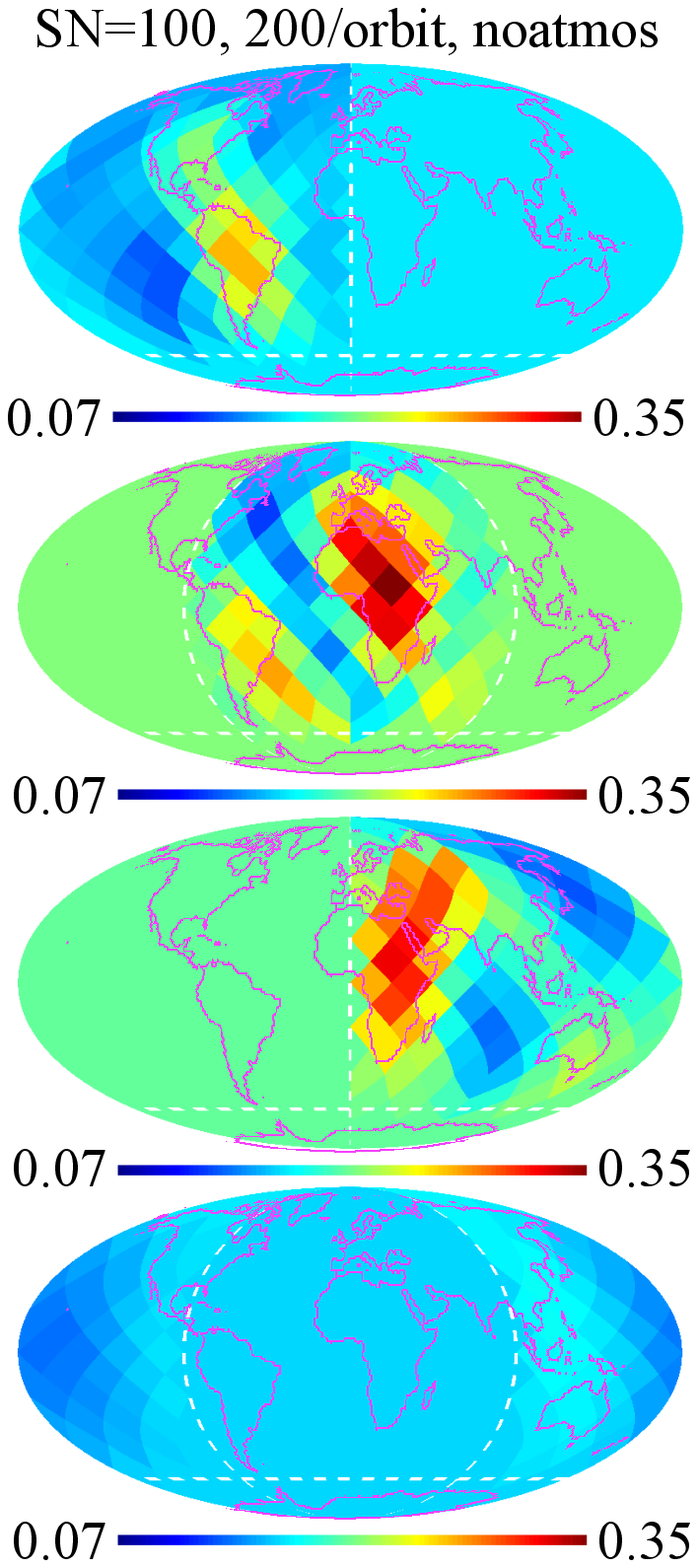}
    \end{center}
  \end{minipage}
  \begin{minipage}{0.484\linewidth}
    \begin{center}
      \includegraphics[width=\linewidth]{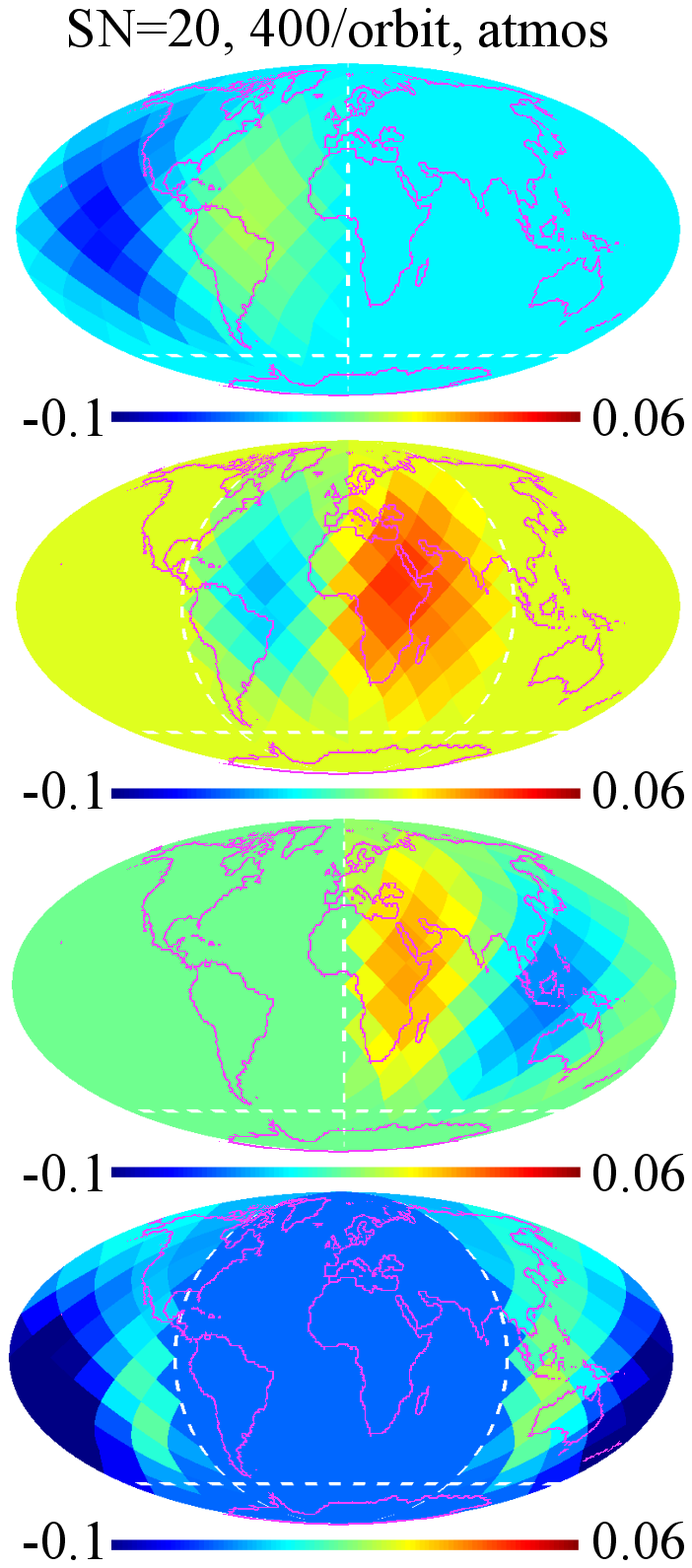}
    \end{center}
  \end{minipage}
    \caption{Left: albedo mapping from NIR band light curve of a hypothetical tidally locked {\it atmosphere-less} Earth. Number of sampling per orbit (with constant intervals) are 200 and all data are used in the analysis ($\Ndata = 200$). 
S/N$=100$ is imposed.  
The white dotted lines are boundaries of illuminated and visible area. 
Right: albedo mapping from NIR$-$Blue light curve of a hypothetical tidally-locked planet with Earth-twin surface and cloud cover. 
Number of sampling per orbit (with constant intervals) are 400 and data at $\alpha \le 90^{\circ }$ are used in the analysis ($\Ndata = 200$). S/N$=20$ is imposed. Assuming S/N$=100$ does not improve the maps significantly. 
}
\label{fig:sync}
\end{figure}

For future projects of direct imaging of HZ planets, those around late-type stars are potentially favorable targets since late-type stars are most abundant in the universe. 
Since HZs around the late-type stars are so close to the star, planets there are likely tidally locked \citep{1993Icar..101..108K}. 
While the illuminated area does not change for the tidally-locked planet, the visible area does change unless a planet is in a face-on orbit (Figure  \ref{fig:gref}(c)), which makes the SOT applicable to such cases as well, though the reconstructable area is inevitably limited to the illuminated half of the surface (see Appendix \appcpdf). 

Here we consider two tidally locked cases (i.e., $\zeta = 0^{\circ }, \omegaspin = \omegaorb $):   
one is an atmosphere-less Earth to see the results for the
idealized case. 
Another is an Earth-twin with cloud cover same as the
previous section. 
In both cases, again, we ignore seasonal variation of snow cover and surface albedos. 
Although those inputs may be unrealistic for actual tidally locked Earth analogs, we avoid getting deeply involved since the aim of this section is to see the basic applicability of the SOT to the tidally locked situation. 

We assume that the sampling frequency is 200/orbit and 400/orbit, respectively, taking into account the fact that the orbit of tidally locked planets is much shorter than Earth's (for instance, the orbital period of an Earth-mass planet around a star with $M = 0.54M_{\odot }$ is 2400 hr). 
Accordingly, we reduce the number of pixels to $\Mmodel = 12\cdot 4\cdot 4 = 192$ ($N_{\rm side}=4)$ in order to avoid underdetermined problems. 

The left column of Figure \ref{fig:sync} displays the recovered maps from the atmosphere-less light curve at the NIR band with S/N$=100$ in the case of $i=60^{\circ }$. 
For demonstrative purpose, we show the maps with $\Phi _{\rm offset} = 0^{\circ }, 90^{\circ }, 180^{\circ }$, and $270^{\circ }$. 
This inversion is in a sense straightforward since our simulation assumes that the surface is Lambertian. 
As expected, the continental distribution is recovered fairly well. 

The right column shows the results of mapping from the NIR$-$Blue light curves of a tidally locked Earth-twin with cloud cover at $\alpha < 90^{\circ }$. 
Though the resolution is now highly deteriorated, there is still inhomogeneity which certainly traces from the real features. 
In this case, increasing S/N to 100 does not improve the image quality significantly.

\subsection{Different Land Distribution}
\label{ss:differentland}

In this section, we consider the sensitivity of reconstruction to the continental distribution. 
One of the ways to do this is to use the continental distribution of ancient Earth \citep[see][]{sanroma2012}. 
We adopt the Earth map in the latest Jurassic period (150 Ma) and that in the early Cambrian period (540 Ma) taken from the paleomap project\footnote{http://www.scotese.com/}\citep{scotese2001} shown by the left column of Figure \ref{fig:lJ}. 
We simply assume that everywhere on the continents has the same soil-like spectrum with reflectivity 0.1 in Blue band and 0.3 in NIR band. 
Since the primary features of cloud pattern are controlled by the global circulation which depends largely on the spin rotation rate and planetary radius, one may expect that global cloud pattern is not completely different in these periods. 
Therefore we simply overlaid clouds of the present Earth on the ancient continental configuration. 

The middle column of Figure \ref{fig:lJ} displays two-dimensional maps retrieved by the same procedure from the NIR$-$Blue light curves with S/N$=20$. The resultant map of the Jurassic Earth shows the primary continental mass concentrated on the center and a smaller feature at the northeast, which trace the input map with lower resolution. Likewise, the map of Cambrian Earth also recovers the input feature. 
The difference among the resultant maps of Earths at different epochs (Figures \ref{fig:map_Earth_inc30_tmu0_sn20} and \ref{fig:lJ}) are evident. 
These results imply that the SOT is applicable to various
continental configurations. 

\begin{figure*}[!tbh]
  \begin{center}
      \includegraphics[width=\hsize]{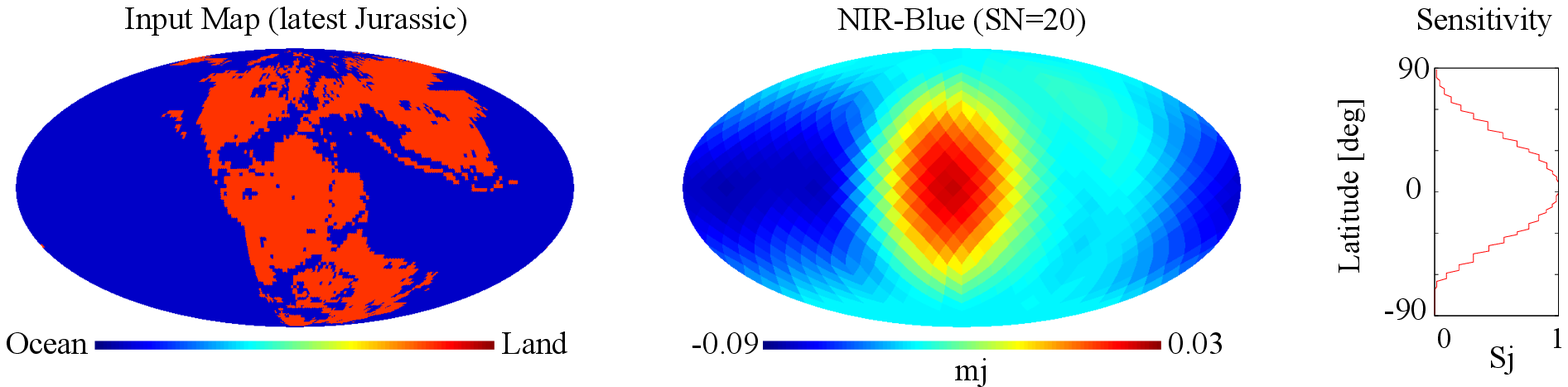}
      \includegraphics[width=\hsize]{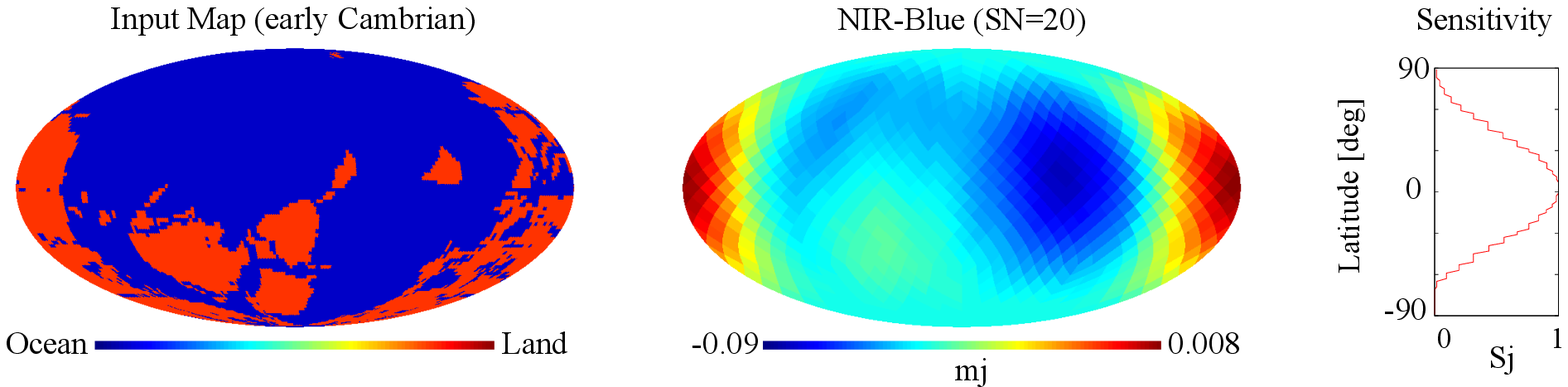}
    \end{center}
    \caption{Left: input data of land distribution based on paleomap of the latest Jurassic period. Right: reconstructed map from NIR$-$Blue light curves with signal-to-noise ratio 20 based on the latest Jurassic (upper panels) and early Cambrian (lower panels) land distribution. Geometric parameters are $\{\zeta , \Theta _{\rm eq}, i\}=\{23^{\circ }.4, 90^{\circ }, 45^{\circ }\}$. Data at $\alpha \le 90^{\circ }$ are used in the analysis. 
The integrated sensitivity $S_j$ is plotted for each case in the right column. }
\label{fig:lJ}
\end{figure*}

\section{Summary and Discussion}
\label{s:sum}

In this paper, we extended the SOT, a method to recover two-dimensional information of exoplanetary surface from annual light curves proposed in Paper I, to be applicable to various geometric cases. 
In Paper I, we considered a planet in a face-on orbit, for which global maps with highly oblique planets are available. 
In this paper, we extend the applicable range of this method to Earth-twins in an arbitrary inclined orbit, which, in some geometric cases, enables the sketch of almost global surface map of low-obliquity planet like the Earth. 
Generally, the effect of anisotropic scattering by cloud can cause bias, but we have shown that omitting the data at crescent phases allows us to reasonably recover the surface inhomogeneity of Earth-twins. 
In the same way as Paper I, an albedo mapping of an Earth-twin from single-band photometry primarily contains the information of the distribution of cloud and snow. 
Information of bare surface under the clouds can be estimated by mapping the difference of light curves of two different photometric bands, thanks to the fact that cloud reflection spectra are roughly independent of wavelengths. 
On the simple assumptions for observational noise described in Section \ref{ss:noise}, a space telescope with aperture size $\sim $ 15 m will allow as to obtain two-dimensional map of an Earth-twin at a distance of 10 pc with satisfactory spatial resolution from continuous 1/2 year observation. 

We tested the reliability of the simultaneous obliquity measurement with mock light curves of Earth-twins with varying directions of the spin axis and the observer. 
We confirmed that the global inhomogeneity of surface albedo allows us to reasonably estimate the planetary obliquity. 

In the sense that primary features in scattered light of the Earth come from clouds because of their global cover (typically more than a half of the entire surface is covered by clouds  with optical thickness larger than 5), clouds can be regarded as a substantial noise source if we aim to know the surface components \citep{ford2001, palle2008, oakley2009, cowan2009, fujii2010, fujii2011}. 
It is even annoying since cloud pattern is variable in the timescale of $\sim $ hour and there is also a seasonal variation. 
Indeed, the short-time variation of cloud cover causes $\sim $10\% fluctuation on diurnal light curves, which is comparable to the observational noise we typically imposed ($\sim $ 12\%).  
Nevertheless, it is known that the annual mean of cloud cover has a characteristic pattern as shown in the bottom panel of Figure \ref{fig:map_Earth_inc30_tmu0_sn20}, which is related to the global atmospheric circulation of the Earth. 
Reconstructed albedo maps are likely to infer such averaged features of cloud cover since the SOT utilizes long-term light curves. 
Therefore we are no longer bothered by temporarily localized cloud cover. 

The SOT has a potential to reveal the presence and rough distributions of the continents. 
For life on the Earth, the presence of continents has played an essential role in evolution of life as a source of nutrients by weathering. 
Nutrients such as phosphorus are indispensable to microbes in ocean such as {\it cyanobacteria}, who has been creating oxygen in Earth's atmosphere. 
Therefore, discovery of continents on HZ planets may encourage us to search for life there. 
In addition, planets with continents can be advantageous targets for the detection of direct indicators of photosynthesis, because the signature of chlorophyll of microbes in ocean is more difficult to detect than the red edge of land plants \citep{2003AsBio...3..531K}. 
The distribution of continents, if exist at all, and the planetary obliquity can greatly affect the clime of the planet \citep[e.g.,][]{williams1997,abe2005}. 
Knowing these properties via the SOT will potentially enable the synthetic discussion of habitability from a variety of aspects.

\acknowledgments

We are thankful to the anonymous referee for many constructive comments. 
We deeply thank Y. Suto, A. Taruya, E. L. Turner, and C. Pichon for useful discussions and comments. 
Y.F. thanks the members of the Department of Astrophysical Sciences of Princeton University for their kind hospitality and suggestive discussions during her visit. 
Y.F. gratefully acknowledges support from the Global Collaborative Research Fund (GCRF) ``A World-wide Investigation of Other Worlds'' grant and the Global Scholars Program of Princeton University. Y.F. and H.K are supported by JSPS (Japan Society for the Promotion of Science) Fellowship for Research, DC:23-6070 and PD:22-5467, respectively.  
We are grateful to OpenCLASTR project for  providing the {\it rstar6b} package.



{\appendix

\section{\appweight. Weight Function}

Here we present the explicit form of the weight function in Equation (\ref{eq:weight}), 
\begin{eqnarray}
W(\phi,\theta;\Phi;\Thetae) &=& \WI(\phi,\theta;\Phi;\Thetae) \, \WV(\phi,\theta;\Phi) \\
\WI(\phi,\theta;\Phi;\Thetae) &=& \mathrm{max}\{  \sin {(\Thetae - \Theta_\mathrm{eq})} \cos {\zeta } \, \sin {\theta } \, \sin {(\phi + \Phi) } + \sin {(\Thetae  - \Theta_\mathrm{eq})} \, \sin {\zeta } \, \cos {\theta } \nonumber \\
&+& \sin {\theta } \, \cos {(\Thetae  - \Theta_\mathrm{eq})} \, \cos {(\phi + \Phi) },0\} \\
\WV(\phi,\theta;\Phi) &=& \mathrm{max}\{ - \sin {i} \, \sin{\Theta_\mathrm{eq}} \, \cos {\zeta } \, \sin {\theta } \, \sin {(\phi + \Phi) - \sin {i} \, \sin{\Theta_\mathrm{eq}} \, \sin {\zeta } \, \cos{\theta }} \nonumber \\
&-& \cos{i} \, \sin {\zeta } \, \sin {\theta } \, \sin {(\phi + \Phi) }+ \cos{i} \cos {\zeta } \, \cos {\theta } + \sin {\theta } \, \sin {i} \, \cos {(\phi + \Phi) } \, \cos {\Theta_\mathrm{eq}}, 0\}.
\end{eqnarray}

The boundary curves of the visible and illuminated area are expressed as 
\begin{eqnarray}
\label{eq:thetaV}
\theta_{V,\pm} (\phi) &=& 2 \tan^{-1} \left( \frac{ a_V(\phi+\Phi) \pm \sqrt{b_V(\phi+\Phi)} }{c_V(\phi+\Phi)} \right) \\
a_V(\phi) &\equiv& - \cos{\zeta} \sin{i} \sin{\phi} \sin {\Theta_\mathrm{eq}} - \sin{\zeta} \cos{i} \sin{\phi} + \sin{i} \cos {\phi} \cos {\Theta_\mathrm{eq}} \nonumber \\
b_V(\phi) &\equiv& \left (- \cos {\zeta} \sin {i} \sin {\phi} \sin {\Theta_\mathrm{eq}} - \sin {\zeta} \cos {i} \sin {\phi}+\sin{i} \cos {\phi} \cos {\Theta_\mathrm{eq}} \right)^2 \nonumber \\
 &+& (\cos {\zeta} \cos {i} - \sin {\zeta} \sin {i} \sin {\Theta_\mathrm{eq}})^2 \nonumber \\
c_V(\phi) &\equiv& \cos {\zeta} \cos {i} - \sin {\zeta} \sin {i} \sin {\Theta_\mathrm{eq}} \nonumber,
\end{eqnarray}
 and 
\begin{eqnarray}
\label{eq:thetaI}
\theta_I(\phi) &=& \left\{
\begin{array}{cc}
\cos^{-1} [- a_I(\phi+\Phi)]  & \mbox{\,\, for $0 \le \Thetae - \Theta_\mathrm{eq} \le \pi$} \\
\cos^{-1} [a_I(\phi+\Phi)] & \mbox{\,\, for $\pi \le \Thetae - \Theta_\mathrm{eq} \le 2 \pi$ } \\
\end{array} \right. \\
a_I(\phi) &\equiv& \displaystyle{ \frac{\cos {\zeta } \sin {(\Thetae - \Theta_\mathrm{eq})} \sin {\phi } + \cos {(\Thetae - \Theta_\mathrm{eq})} \cos {\phi }}{\sqrt{ \sin ^2{(\Thetae - \Theta_\mathrm{eq})}  \sin^2 {\zeta} + \left( \cos (\Thetae - \Theta_\mathrm{eq}) \cos \phi + \cos \zeta  \sin (\Thetae - \Theta_\mathrm{eq}) \sin \phi \right)^2 }} }\nonumber.
\end{eqnarray}

Using the spherical coordinate $(\phiobs, \thetaobs)$ to the observer on the planetary surface, one can rewrite Equation (\ref{eq:thetaV}) to
\begin{eqnarray}
\theta_{V,\pm} (\phi) = \cos^{-1} \left(\pm \frac{\cos{(\phi-\phiobs)} \sin{\thetaobs}}{\sqrt{\cos^2{\thetaobs} + \cos^2{(\phi - \phiobs)} \sin^2{\thetaobs} }} \right).
\end{eqnarray}
We note that $\phiobs = 2 \pi - \Phi$ and $\theta_{V,\pm} (\phi)$ does not depend on $\Theta$.

Converting $\elos$ to the spherical surface at $\Phi=0$ (i.e. $R_x(-\zeta) \elos$) , we obtain the transformation formula to $(\phiobs. \thetaobs)$ as
\begin{eqnarray}
\cos{\thetaobs} &=& \ci \czeta - \si \szeta \sin{\Theta_\mathrm{eq}} \\
\sin{\thetaobs} \sin{{(\phiobs+\Phi)}} &=& - \ci \szeta - \si \czeta \sin{\Theta_\mathrm{eq}} \\
\sin{\thetaobs} \cos{{(\phiobs+\Phi)}} &=& \si \cos{\Theta_\mathrm{eq}}.
\end{eqnarray}

\section{\appcpdf. Reconstructable Area}

\begin{figure}[!tbh]
  \centerline{\includegraphics[width=150mm]{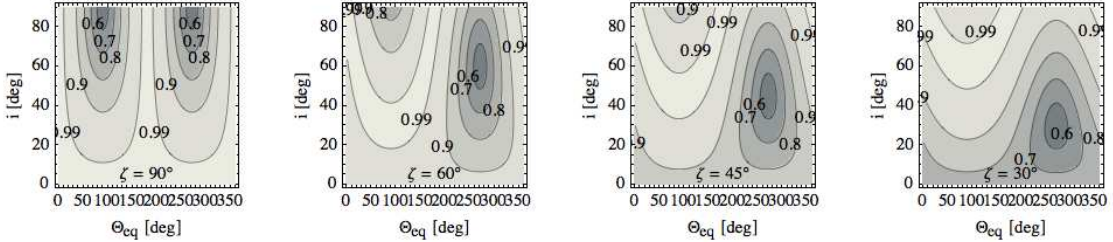}}
  \caption{Contour plot of the cover fraction of the illuminated and visible surface over a planetary year as a function of $i$ and $\Theta _{\rm eq}$. Solid contours show $\fc = 0.6, 0.7,0.8,0.9,$ and $0.99$. Each panel displays different obliquity corresponding to $\zeta = 90^\circ, 60^\circ, 45^\circ$, and $30^\circ$. \label{fig:cover}}
\end{figure}

Figure \ref{fig:cover} shows the contour of the area cover fraction of the surface sampled at least once during a planetary year $\fc$, that is the maximum reconstructable area,  for various geometric configurations. We also derive the cumulative probability distribution function $p(>\fc)$. The whole region of the visible area is covered in a planetary year, we only have to consider the cover fraction of the visible area ($p(>\fc)$ does not depend on the obliquity). Considering diurnal rotation, we can express the cover fraction as a function of $\thetaobs$,
\begin{eqnarray}
\fc = \frac{1 + \sin{\thetaobs}}{2}
\label{eq:fcobs}
\end{eqnarray}
Using the uniform distribution of the line of sight $p(\fc) \, d \fc = \sin{\thetaobs} \, d \thetaobs$ and Equation (\ref{eq:fcobs}), we obtain
\begin{eqnarray}
p(>\fc) &\equiv& \int_{\fc}^1 p(\fc) d \fc \nonumber =\int_{\fc}^1 \sin{\thetaobs} \left| \frac{d \thetaobs}{d \fc} \right| d \fc \nonumber \\
 &=& \int_{\fc}^1 \frac{2 (2 \fc -1)}{\sqrt{1-(2 \fc -1)^2}} d \fc = 2 \sqrt{(1-\fc) \fc}.
\end{eqnarray}
The $p(>\fc)$ does not depend on the obliquity $\zeta$ and one can get above 95\% (80\%) of the surface information in 43.5\% (80\%) for the randomly selected exoplanets, in principle. 

When a planet is in a synchronous orbit, a half of the planetary surface is always dark. We obtain
\begin{eqnarray}
\fc = \frac{1 + \sin{i}}{4}.
\label{eq:fcobst}
\end{eqnarray}
The CPDF of the tidally locked planet is expressed as  
\begin{eqnarray}
p_{\mathrm{TL}}(>\fc) &=& \int_{\fc}^{\frac{1}{2}} \sin{i} \left| \frac{d i}{d \fc} \right| d \fc \nonumber \\
 &=& \int_{\fc}^{\frac{1}{2}} \frac{\sqrt{2} (-1 + 4 \fc)}{\sqrt{\fc (1 - 2 \fc)}} d \fc = 2 \sqrt{2 (1- 2 \fc) \fc}.
\end{eqnarray}
Both $p(>f_{\rm IV})$ \& $p_{\rm TL}(>f_{\rm IV})$ are shown in Figure \ref{fig:covercum}.

\begin{figure}[!tbh]
  \centerline{\includegraphics[width=80mm]{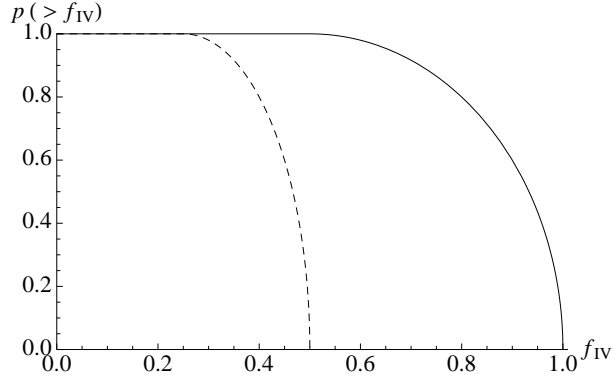}}
  \caption{Cumulative probability distribution function (CPDF) of the cover fraction $\fc$ on the assumption of the random line of sight ($p(i,\Omega) \propto \sin i $), $p(>\fc) = 2 \sqrt{\fc (1-\fc)}$. This function is independent of the obliquity $\zeta$. The probability $p(\fc > 95\%)$ is 43.5\% and $p(\fc > 80\%)$ is 80\%. The dashed line indicates the CPDF for tidally locked planets: $p_{\mathrm{TL}}(>\fc) = 2 \sqrt{2 \fc (1- 2 \fc) }$. The probability $p_{\mathrm{TL}}(\fc > 47.5\%)$ is 43.5\% and $p_{\mathrm{TL}}(\fc > 40\%)$ is 80\%. \label{fig:covercum}}
\end{figure}

\section{\appbayes. Bayesian Interpretation of the Tikhonov Regularization}

Following \citet{tarantola2005}, we discuss the Bayesian interpretation of the Tikhonov regularization. 
The probability of a posterior information in the model space is given by
\begin{eqnarray}
p_M(\mv) \propto \rho_M(\mv) \rho_D(\gv (\mv)), 
\end{eqnarray}
Here we neglect modelization uncertainties. Introducing a Gaussian model for the a priori information on the model parameters with the covariance matrix $\Cm$,
\begin{eqnarray}
\rho_M(\mv) = \frac{1}{\sqrt{(2 \pi)^\Mmodel} \det{\Cm}} \exp{\left[ -\frac{1}{2} (\mv - \mpr)^\trans \Cm^{-1} (\mv - \mpr) \right]}, 
\end{eqnarray}
and assuming the statistical noise is an independent Gaussian noise with the covariance matrix $\Cd$
\begin{eqnarray}
\rho_D(\dv) = \frac{1}{\sqrt{(2 \pi)^\Ndata} \det{\Cd}} \exp{\left[ -\frac{1}{2} \errorv^\trans \Cd^{-1} \errorv \right]},
\end{eqnarray}
one can obtain
\begin{eqnarray}
P_M(\mv) \propto \exp{ \left\{ -\frac{1}{2} \left[ \errorv^\trans \Cd^{-1} \errorv  + (\mv - \mpr)^\trans \Cm^{-1} (\mv - \mpr) \right] \right\}} .
\end{eqnarray}
The model which maximizes a posteriori probability is given by minimizing 
\begin{eqnarray}
Q = \errorv^\trans \Cd^{-1} \errorv  + (\mv - \mpr)^\trans \Cm^{-1} (\mv - \mpr) .
\end{eqnarray}
If taking the independent Gaussian noise $(\Cd)_{ij} = \sigma_i^2 \delta_{ij} $ and  $(\Cm)_{ij} = \lambdar^{-2} \delta_{ij} $,  we obtain the Tikhonov regularization , 
\begin{eqnarray}
Q = \sum \frac{|d_i - g_i(\mv)|^2}{\sigma_i^2}  + \lambdar^2 |\mv - \mpr|^2.
\label{eq:qminimization_MAP}
\end{eqnarray}

\section{\applcurve. The L-curve Criterion}
\label{ap:lcurve}

\begin{figure}[!tbh]
  \begin{minipage}{0.50\linewidth}
    \begin{center}
      \includegraphics[width=\linewidth]{fig14a.ps}
    \end{center}
  \end{minipage}
  \begin{minipage}{0.30\linewidth}
    \begin{center}
      \includegraphics[width=\linewidth]{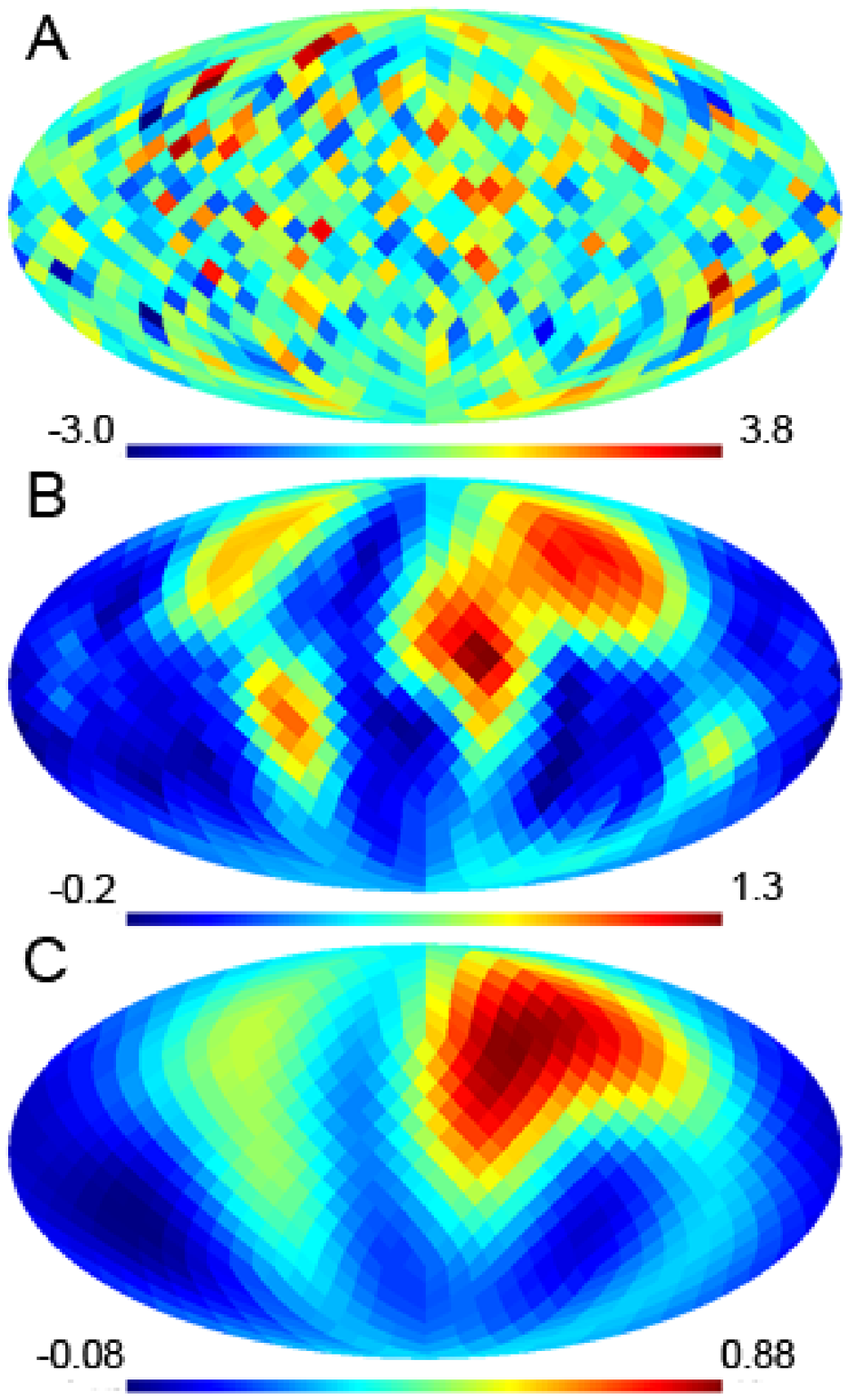}
    \end{center}
  \end{minipage}
 \caption{$\lambdar$ dependence of the best-fit model. The left panel shows the L-curve for the mock observations with $\zeta=90^\circ,\Theta _{\rm eq}=0^\circ,i=0^\circ$, and the additional 5\% noise. The input data are a land/ocean binary map of the Earth, which has 1 on land and 0 on ocean. A solid curve indicates the parametric plot of $|\tG {\mestl} - \tdv|$ and $|\mestl - \mpr|$  as a function of the regularization parameter $\lambdar$. 
 A dotted curve shows the curvature of the L-curve, $\curv$. The definition of $\curv$ is written in Appendix \appcurv. 
 We took three different regularization parameters: $\lambdar=10^{-0.5}$ (A), $10^{0.48}$, which corresponds to the maximum curvature (B), and $\lambdar=10^{1.5}$ (C). The recovered models are displayed in the right panels: the upper, middle, and lower panels correspond to A, B, and C.\label{fig:lam}}
\end{figure}

The regularization parameter $\lambdar$ is chosen based on ``the L-curve criterion'' following \citet{hansen2010}. 
The L-curve is a parametric plot of the model norm $|\mestl-\mpr|$ and the residuals $|\tG \, \mestl - \tdv|$  as a parametric function of $\lambdar$. 
In the left panel of Figure \ref{fig:lam}, we demonstrate the L-curve of the mock observations of an atmosphere-less Earth with $\zeta=90^\circ,\Theta _{\rm eq}=0^\circ,i=0^\circ $ and 3\% noise. 
As $\lambdar$ increases, $|\mestl-\mpr |$ rapidly decreases until {\it B}  (the L-curve's corner), and then the slope becomes gentler. 
The large value of $|\mestl-\mpr|$ (small $|\tG \, \mestl - \tdv|$) implies that the perturbation error due to observational noise is large, while large $|\tG \, \mestl - \tdv|$ (small $|\mestl-\mpr|$), indicates that the bias due to the regularization (regularization error) is large. 
The L-curve corner serves as the point of good balance between these errors. 
The L-curve corner is defined by the maximum curvature in the log-log space, as described in Appendix \appcurv \citep{hansen2010}. Adopting different values for $\lambdar $ to Equation (\ref{eq:tikres}), we see the $\lambdar$ dependence of the recovered models $\mestl$ in the right panels. As shown in the left panel, the L-curve criterion gives $\lambdar=10^{0.48}$ at the maximum curvature point (B). 
Right panel of  Figure \ref{fig:lam} shows the recovered maps with $\lambdar=10^{-0.5}$ (A),  the L-curve corner $\lambdar=10^{0.48}$ (B), and $\lambdar=10^{1.5}$ (C). 
The recovered map for A exhibits large perturbation noise, while the resolution of the model is degraded for the case (C). 
The L-curve criterion balances these two opposite effects (B).

\section{\appcurv. Maximum curvature of the L-curve}

Here, we briefly describe the derivation of the curvature of the L-curve following \citet{hansen2010}.  We first introduce $\xi \equiv |\mestl-\mpr|^2$ and $\rho \equiv |\tG \, \mestl - \tdv|^2$. The curvature of the L-curve is defined in the $\log{|\tG \, \mestl - \tdv|}-\log{|\mestl-\mpr|}$ space, 
\begin{eqnarray}
\curv \equiv - 2 \frac{\ldrho \lddxi - \lddrho \ldxi}{[\ldrhot +\ldxit]^{3/2}},
\label{eq:curvdef}
\end{eqnarray}
where the prime indicates the derivative by $\lambdar$. The derivatives of $\xi \equiv |\mestl-\mpr|^2$ and $\rho = |\tG \, \mestl - \tdv|^2$ are
\begin{eqnarray}
\label{eq:adxi}
\dxi &=& -\frac{4}{\lambdar} \sum_{i=1}^{\Mmodel} [1-w_i(\lambdar)] w_i^2(\lambdar) \frac{\Delta _i^2}{\eig_i^2} \\
\label{eq:adrho}
\Delta _i^2 &\equiv& \sum_{j=1}^{\Ndata }\left[ u_{ji} \left(\tilde d_j-\sum _{k}^{\Mmodel} \tilde G_{jk} \mpr_k \right) \right]^2 \\
\drho &=& - \lambdar^2 \dxi \\
w_i(\lambdar) &\equiv& \displaystyle{\frac{\eig_i^2}{\eig_i^2 + \lambdar^2}}.
\end{eqnarray}
Using Equations (\ref{eq:adxi}) and (\ref{eq:adxi}), one obtain
\begin{eqnarray}
\curv = - 2 \frac{\xi \rho }{\dxi} \frac{(\lambdar^2 \dxi \rho + 2 \lambdar \xi \rho + \lambdar^4 \xi \dxi)}{(\lambdar^4 \xi^2 + \rho^2)^{3/2}}.
\label{eq:curvdefd}
\end{eqnarray}

}

\end{document}